\documentclass{aa}  

\usepackage{graphicx}
\usepackage{txfonts}
\begin{document}

   \title{The reduced proper motion selected halo:\\methods and description of the catalogue\thanks{The full catalogue is available in electronic form
at the CDS via anonymous ftp to \protect\url{cdsarc.u-strasbg.fr} (130.79.128.5)
or via \protect\url{http://cdsweb.u-strasbg.fr/cgi-bin/qcat?J/A+A/}}}
   \titlerunning{The RPM halo: methods and description of the catalogue}
   \author{Helmer H. Koppelman
          \and
          Amina Helmi}
   \authorrunning{H.H. Koppelman \and A. Helmi}
   \institute{Kapteyn Astronomical Institute, University of Groningen, Landleven 12, 9747 AD Groningen, The Netherlands\\
              \email{koppelman@astro.rug.nl}
}
   \date{}

  \abstract
   {The {\it Gaia} mission has provided the largest ever astrometric chart of the Milky Way. Using it to map the Galactic halo is helpful for disentangling its merger history.
  }
  {
 The identification of halo stars in {\it Gaia} DR2 with reliable distance estimates requires special methods because such stars are typically farther away and scarce.
  }
  {
   We applied the reduced proper motion (RPM) method to identify halo main sequence stars on the basis of {\it Gaia} photometry and proper motions. Using the colour-absolute-magnitude relation for this type of star, we calculated photometric distances. Our selection results in a set of $\sim10^7$ tentative main sequence halo stars with typical distance
uncertainties of $7\%$ and with median velocity errors of 20~km/s. The median distance of our sample is $\sim 4.4$~kpc, with the faintest stars located at $\sim 16$~kpc. 
  }
  {
  The spatial distribution of the stars in our sample is centrally concentrated. A visual inspection of the mean velocities of stars on the sky reveals large-scale patterns as well as 
clear imprints of the GD-1 stream and tentative hints of the Jhelum and Leiptr streams. Incompleteness and selection effects limit our ability to interpret the patterns reliably as well as to identify new substructures. We define a pseudo-velocity space by setting the line-of-sight velocities of our sample stars to zero. In this space, we recover 
several known structures such as the footprint of Gaia-Enceladus (i.e. the Gaia-Sausage) as well as the Helmi Streams and some other retrograde substructures (Sequoia, Thamnos). 
   We show that the two-point velocity correlation function reveals significant clustering on scales smaller than 100~km/s of a similar amplitude as found for the 6D {\it Gaia} halo sample. This clumping of stars in velocity space might hint at the presence of nearby streams that are predominantly phase-mixed.
  }
  {
  A spectroscopic follow-up of our halo main sequence sample is bound to yield unprecedented views of Galactic history and dynamics. In future {\it Gaia} data releases, the level of systematics will be reduced and the astrometry will be more precise, which will allow for the identification of more substructures at larger distances.}


   \maketitle
%

\section{Introduction}
Galactic cartography has gained a huge boost with the advent of astrometric, photometric, and spectroscopic data of unprecedented volume and quality from the {\it Gaia} mission \citep{GaiaCollaboration2016TheMission, GaiaCollaboration2018brown}, in combination with spectroscopic surveys such as APOGEE \citep{Wilson2010TheSpectrograph,Abolfathi2018TheExperiment}, RAVE \citep{Kunder2017}, and LAMOST \citep{Cui2012}. These datasets have revealed numerous substructures in different Galactic components. For example, large wave-like patterns and sharp ridges have been found in the stellar disc \citep{Antoja2018WrinklesDisk, Kawata2018RadialGaiaDR2, Katz2018MappingKinematics}. These arches give rise to intricate structures in action-space \citep{Trick2019TheDR2}, and their imprint varies with location in the disc \citep{Ramos2018RidingGaia}. They are likely due to a combination of perturbations of a satellite flying by \citep{Minchev2009IsStreams, Antoja2018WrinklesDisk, Laporte2018FootprintsSet} and internal dynamical processes such as resonances with the bar and spiral structures \citep{Monari2019SignaturesSpace, Hunt2019SignaturesDisc, Khanna2019TheWay, Chiba2019ResonanceBar}. 

Studies, such as those mentioned above, show how complex and intertwined the phase-space structure of the Milky Way is. The expectation has been that the Galactic halo would be similarly complex as a result of the mergers experienced by the Milky Way \citep{Helmi1999a}. In fact, the analysis of {\it Gaia} DR2 has revealed that besides several small features, the local stellar halo is largely dominated by two structures, including a large, radially-anisotropic, slightly retrograde kinematic structure and a hot thick disc \citep{Belokurov2018Co-formationHalo, Koppelman2018, Haywood2018InDR2}. The origin of the former is related to the accretion of a massive $(M_\star\sim10^{9}~{\rm M}_\odot)$ dwarf galaxy known as Gaia-Enceladus \citep[or Gaia-Sausage, c.f. \citealt{Belokurov2018Co-formationHalo}]{Helmi2018}. The accretion of Gaia-Enceladus took place 9-11 Gyr ago \citep{Helmi2018, DiMatteo2018, Mackereth2019TheSimulations, Chaplin2020AgeIndi} and lead to heating of the proto-Milky Way disc \citep{Helmi2018, Gallart2019UncoveringGaia, Chaplin2020AgeIndi}. 

A large fraction of the results listed above stem from the {\it Gaia} sub-sample containing full velocity information \citep{Katz2019GaiaVelocities}, also known as the 6D sample. Although impressive in size, the 6D sample is small relative to the 5D sample (i.e. without line-of-sight velocities). It comprises $\sim 7$ million sources compared to a staggering $\sim 1.3$ billion sources with parallaxes and proper motions. Most stars in {\it Gaia} DR2 have relatively large errors in their parallax measurements, as `only' $\sim 150$ million sources have distances with relative errors $<20\%$. Also the {\it Gaia} DR2 parallaxes are known to suffer from a zero-point offset of $\sim -0.029~{\rm mas}$ with a root mean square (RMS) of $0.03-0.05~{\rm mas}$, which varies with the location on the sky \citep{Arenou2018, GaiaCollaboration2018brown, Lindegren2018}. For brighter stars, the typical off-set may be closer to $\sim 0.05~{\rm mas}$ \citep{Schonrich2019, Leung2019SimultaneousLearning, Zinn2019ConfirmationField, Chan2019TheStars}. The poorer parallaxes in combination with the missing line-of-sight velocities therefore complicate the utilisation of the entire {\it Gaia} DR2 sample for dynamical studies.

A general approach to cope with poorly constrained or missing distances is to use the luminosity of the sources that are known as `standard candles', such as RR-Lyrae. Their period-luminosity relation can be used to derive distances that are typically accurate up to $\sim 5\%$. This makes RR-Lyrae outstanding targets to study the morphology of the disc and stellar halo \citep[e.g.][]{Watkins2009, Sesar2010, Drake2012, Hernitschek2018, Iorio2019TheMerger, Zinn2019LocalAlien}. The downside of RR-Lyrae stars is that they are not abundant. Another approach is to identify stars of a specific type such as blue horizontal-branch (BHB) stars \citep{Xue2008, Xue2011QuantifyingHalo, Deason2012TheHalo, Fukushima2017StructureStars, Lancaster2019TheStars}, or red-giant branch (RGB) stars \citep{Morrison2009FASHIONABLYHALO} whose absolute magnitude can be derived using isochrones when knowledge of their surface gravity $(\log{g})$ or metallicity is available, such as from spectroscopic surveys \citep[e.g.][]{ Leung2019SimultaneousLearning, Conroy2019MappingSurvey, Cargile2019MINESweeper:Era}.

In this paper, we use main sequence (MS) stars to study the Milky Way halo. The reason for focusing on MS stars is twofold. Firstly, they follow a relatively simple absolute magnitude relation as a function of colour, which can be used to calculate a photometric distance \citep[e.g.][]{Juric2008THEDISTRIBUTION, Ivezic2008TheMetallicity, Bonaca2012UPDATEOVERDENSITY}. 
Secondly, we can select them using only their photometry and proper motions (i.e. without knowing the distance to the stars) through a property known as the reduced proper motion \citep[e.g.][]{Jones1972REDUCED-PROPER-MOTIONDIAGRAMS, Smith2009}.

We describe the methods that we use in Sect.~\ref{sec:methods}, and the selection of MS halo stars and calibration of their distances in Sect.~\ref{sec:data}. In Sect.~\ref{sec:spatdist} we explore the spatial distribution of the halo stars in the sample. We combine the spatial coordinates with the proper motions of the stars to calculate tangential velocities. In Sect.~\ref{sec:veldist} we inspect the velocities of the full sample and in Sect.~\ref{sec:velloc} we focus on the velocity distribution of a local sample. Finally, in Sect.~\ref{sec:conclusions} we summarise and discuss our results, and present our conclusions.

\section{Methods}\label{sec:methods}
In this section, we describe the tools that we require to select halo MS stars and to calibrate photometric distances. The description of the data, selection, and calibration will be carried out in Sect~\ref{sec:data}.

\subsection{Photometric distance estimates for MS stars}

Distances are notoriously difficult to measure in astronomy. Only about $\sim 10\%$ of the parallaxes released in {\it Gaia} DR2 are precise (i.e. those with  ${\tt parallax\_over\_error}>5$). Another way of calculating distances is through the luminosity of a star. For specific types of stars, for which the intrinsic luminosity is known, we can calculate a photometric distance from the apparent luminosity. The relation between the intrinsic and apparent magnitude of a star in the {\it Gaia} $G$-band is given by
\begin{equation}
    M_G = m_G - 5\log_{10}(\frac{d}{\rm kpc}) - 10 - A_G,
    \label{eq:absmag}
\end{equation}
where $M_G$ is the absolute magnitude of the star, $m_G$ is its apparent magnitude, $d$ is its distance, and $A_G$ is the extinction in the $G$-band. For most sources, $M_G$ is unknown and Eq.~\eqref{eq:absmag} cannot be used to calculate a distance $d$. 

In this work, we use the close to linear relation of the colour and absolute magnitude of MS stars to derive a distance. We note that the MS is only approximately linear in optical pass-bands, and in the near-infrared this approximation breaks down \citep[e.g. Fig.~9 of][]{GaiaCollaboration2018Babusiaux}. Because of the strong correlation between $M_G$ and colour of the MS in the Hertzsprung-Russel diagram (HRD) we can find the distance independent relation
\begin{equation}
    M_G = f(G-G_{\rm RP}).
    \label{eq:absfcol}
\end{equation}
We use the {\it Gaia} ${G- G_{\rm RP}}$ colour because it is less prone to systematic effects than ${ G_{\rm BP}- G_{\rm RP}}$, especially in crowded fields \citep{GaiaCollaboration2018brown}. Because ${G- G_{\rm RP}}$ is distance independent, we can use Eq.~\eqref{eq:absmag} to calculate a distance that is a function of the apparent magnitude and colour only
\begin{equation}
    d = 10^{(m_G - M_G - 10 - A_G)/5}.
    \label{eq:dist}
\end{equation}
When propagating the error in $d$, assuming that the error in $m_G$ can be neglected, we find that the relative distance error is
\begin{equation}
    \epsilon_d/d = 0.2\log(10) \epsilon_{M_G},
    \label{eq:edist}
\end{equation}
where $\epsilon_{M_G}$ is the error in the absolute magnitude.

\subsection{Selecting MS stars}
The method of determining distances that is described above is valid for MS stars. Giants and stars at the MS turn-off (MSTO) describe a sequence in the HRD that is too vertical or degenerate to find a reliable relation between $M_G$ and colour. Therefore, we need a (distance independent) way of selecting MS stars. 

We identify MS stars using a combination of the proper motion and photometry known as a reduced proper motion \citep[RPM, ][see also \citealt{Smith2009}]{Jones1972REDUCED-PROPER-MOTIONDIAGRAMS}. The RPM of a star, in the {\it Gaia} G-band, is defined as
\begin{equation}\label{eq:Hg}
    H_G \equiv m_G + 5\log_{10}(\frac{\mu}{\rm mas/yr}) - 10 - A_G,
\end{equation}
where $\mu = \sqrt{\mu_\ell^2+\mu_b^2}$ is the amplitude of the proper motion. This equation is similar to Eq.~\eqref{eq:absmag} and the two can be combined to gain some insight
\begin{equation}
    H_G = M_G + 5\log_{10}(\frac{v_\mathrm{tan}}{4.74057 ~{\rm km/s}}),
    \label{eq:Hg2}
\end{equation}
where $v_\mathrm{tan}$ is the tangential velocity of a star given by
\begin{equation}\label{eq:vtan}
    v_\mathrm{tan} = 4.74057~{\rm km/s}~
    \bigg(\frac{\mu}{\rm mas/yr}\bigg)~
    \bigg(\frac{d}{\rm kpc}\bigg),
\end{equation}
where $d$ is the heliocentric distance to the star.

When plotted as a function of colour, the $H_G -$colour diagram (which we refer to as the RPM diagram) of a stellar population is equal to the HRD — but with an offset that depends on $v_\mathrm{tan}$. If all the stars in the population have the same $v_\mathrm{tan}$, their sequence in RPM diagram and HRD will look exactly the same. However, if the stellar population has a mean $v_\mathrm{tan}$ plus a few km/s dispersion, its sequence in the RPM diagram will be broadened by the logarithm of the velocity dispersion. Furthermore, populations with characteristic, specific tangential velocities will split into parallel sequences.

We exploit this splitting of the MS to select halo stars by identifying the region where the MS stars with high $v_\mathrm{tan}$ are located. Since the $v_\mathrm{tan}$ for the disc is small, even when considering the dispersion, the halo should appear as a separate sequence. Eqs.~\eqref{eq:Hg} and \eqref{eq:Hg2} imply that, for fixed $v_\mathrm{tan}$ populations, $H_G$ will only be a function of ${G- G_{\rm RP}}$ and can readily be computed. At the core of our selection method, we aim to locate high-$v_\mathrm{tan}$, MS stars in the RPM diagram.

This type of selection is conceptually not too different from a kinematic selection performed in the Toomre diagram, which is often used to identify halo stars in the 6D sample of {\it Gaia} \citep[e.g.][]{Nissen2010, Bonaca2017, Posti2018TheEllipsoid, Koppelman2018}. Reduced proper motion diagrams are often used to select white dwarfs and classify them as belonging to the halo or the disc \citep[e.g.][]{Kalirai2004Thesub0/sub, Kilic2006CoolSurvey, Fusillo2015A10, Torres2019RandomPopulation, Geier2020Population2}. However, it is important to note that the RPM selection has a clear bias: halo stars with a small tangential velocity will not be selected (e.g. halo stars moving only along the line-of-sight).
\section{Data selection and calibration}\label{sec:data}

\subsection{Data and quality cuts}\label{sec:photqual}
We start from the full sub-sample of {\it Gaia} DR2 with multi-band photometry. For the method outlined in Sect.~\ref{sec:methods}, we have to rely on the photometry of the sources. Therefore, we impose the several cuts on the photometric quality of the stars. Firstly, we select stars with {\tt phot\_g\_mean\_flux\_over\_error} > 50 and {\tt phot\_rp\_mean\_flux\_over\_error} > 20. Secondly we require stars to have {\tt phot\_bp\_rp\_excess\_factor} < 1.3 + 0.06$\cdot ({\tt bp\_rp})^2$ and {\tt phot\_bp\_rp\_excess\_factor} > 1.0 + 0.015$\cdot ({\tt bp\_rp})^2$,
where ${\tt bp\_rp} = {\tt phot\_bp\_mean\_mag} - {\tt phot\_rp\_mean\_mag}$. Besides cleaning the photometry, these cuts also remove sources with a bad astrometric solution \citep[see][]{Arenou2018}. 

We use the re-normalised unit weight error (RUWE) to further clean the sample. When DR2 came out, the {\it Gaia} Data Processing and Analysis Consortium (DPAC) recommended using the unit weight error (UWE) to filter sources with a bad astrometric solution \citep[e.g.][]{Lindegren2018, Arenou2018}. However, the original UWE varies with colour and magnitude. Therefore, DPAC \citep{Lindegren2018b} recommends the use of the re-normalised UWE (RUWE), which does not depend on stellar properties. Following their suggestion, we remove all the stars that have RUWE $>1.4$.

Independently of the quality of the photometry, our sources are prone to extinction caused by absorption from dust in the interstellar medium (ISM) along the line-of-sight. We correct for the extinction using the \cite{Schlegel1998MAPSFOREGROUNDS} maps. 

These maps provide the extinction factor integrated along the entire line-of-sight. For distant stars we assume that all the absorbing ISM clouds lie in the foreground. However, for nearby stars we have to be careful, as some of the extinction in the \citeauthor{Schlegel1998MAPSFOREGROUNDS} maps might come from regions in the ISM behind the stars. Therefore, we use the approach outlined by Eqs.~(10) \& (11) from \cite{Binney2014NewStars} to calculate the amount of foreground dust for each star as a function of its parallax and the location on the sky. The amount of extinction is given by 
\begin{equation}
A_V(\ell,b,s) = A_{V,\infty}(\ell,b)
\frac
{\int_0^s \rho[\boldsymbol{x}(s')] {\rm d}s'}
{\int_0^\infty \rho[\boldsymbol{x}(s')] {\rm d}s'},
\end{equation}
where $A_{V,\infty}(\ell,b)$ is the extinction given by the \citeauthor{Schlegel1998MAPSFOREGROUNDS} maps, $s$ is the heliocentric distance to the star, and $\boldsymbol{x}$ is the position vector of the star that lies at distance $s$ in the direction of $(\ell,b)$ on the sky. For the dust density we follow the model presented by Eq.~(16) of \cite{Sharma2011GALAXIA:WAY} 

\begin{equation}
    \rho_{\rm Dust}(R,z) = \exp{\bigg(\frac{R_\odot-R}{h_R}} 
    {-\frac{|z-z_{\rm warp}|}{k_{\rm flare}h_z}\bigg)}
\end{equation}
where $z_{\rm warp}$ and $k_{\rm flare}$ describe the warping and flaring of the disc
\begin{equation}
    k_{\rm flare}(R) = 1+\gamma_{\rm flare}{\rm Min}(R_{\rm flare},R-R_{\rm flare})
\end{equation}
and
\begin{equation}
    z_{\rm warp}(R,\phi) = \gamma_{\rm warp}{\rm Min}(R_{\rm warp},R-R_{\rm warp})\sin{(\phi)}
\end{equation}
with values $h_R = 4.2~{\rm kpc}$, $h_z = 0.088~{\rm kpc}$, $\gamma_{\rm warp} = 0.18~{\rm kpc^{-1}}$, $R_{\rm warp} = 8.4~{\rm kpc}$, $R_{\rm flare} = 1.12~R_\odot$, and $\gamma_{\rm flare} = 0.0054~{\rm kpc^{-1}}$, which are based on the model of \cite{Robin2003AWay}.
Effectively, we derive in this way an extinction fraction $\frac{A_V(b,\ell,s)}{A_{V,\infty}(b,\ell)}$ which encodes what fraction of the full extinction correction should be applied.

Following \cite{Binney2014NewStars}, we scale the \citeauthor{Schlegel1998MAPSFOREGROUNDS} maps because the reddening in the regions $E(B-V)>0.15$ is overestimated \citep[e.g.][]{Arce1999MeasuringTest}. The correction factor that we apply is 
\begin{equation}
    f(E(B-V)) = 0.6 + 0.2\bigg[1-\tanh{\bigg(\frac{E(B-V)-0.15)}{0.3}}\bigg)\bigg].
\end{equation}
This factor scales the highly reddened regions by a factor of $0.6$. We note that the results that are presented in this work are not affected by this scaling.

For stars with parallaxes smaller than $ 0.1~{\rm mas}$ we apply the full extinction fraction because they are likely to be distant stars. For all the other stars we invert the parallaxes to obtain an estimate for the distance. We ignore the error on the parallax because we only are looking for an estimate of the distance. On average, the parallax-errors increase with heliocentric distance. Nearby sources, for which the correction fraction is essential, will have relatively good parallaxes. The correction fraction is larger than $0.90$ for more than $90$\% of the sources, only $1.6$\% of the stars receive a correction of smaller than $0.50$.
The resulting weighted $A_V$ values are transformed to $A_G$, $A_{BP}$, and $A_{RP}$ using the relations given by \cite{Malhan2018GhostlyCatalogue} \citep[they originate from the Padova model\footnote{http://stev.oapd.inaf.it/cmd} and are originally based on][]{Cardelli1989TheExtinction}.

As a final quality selection criterion we remove sources located in areas on the sky where the extinction is larger than $A_V>2.0$. These highly extinct sources are mostly found close to the plane of the disc, so the cut acts as a filter for the Galactic disc.

\subsection{Fitting the MS}\label{sec:MSfit}
\begin{figure}
    \centering
    \includegraphics[width=\hsize]{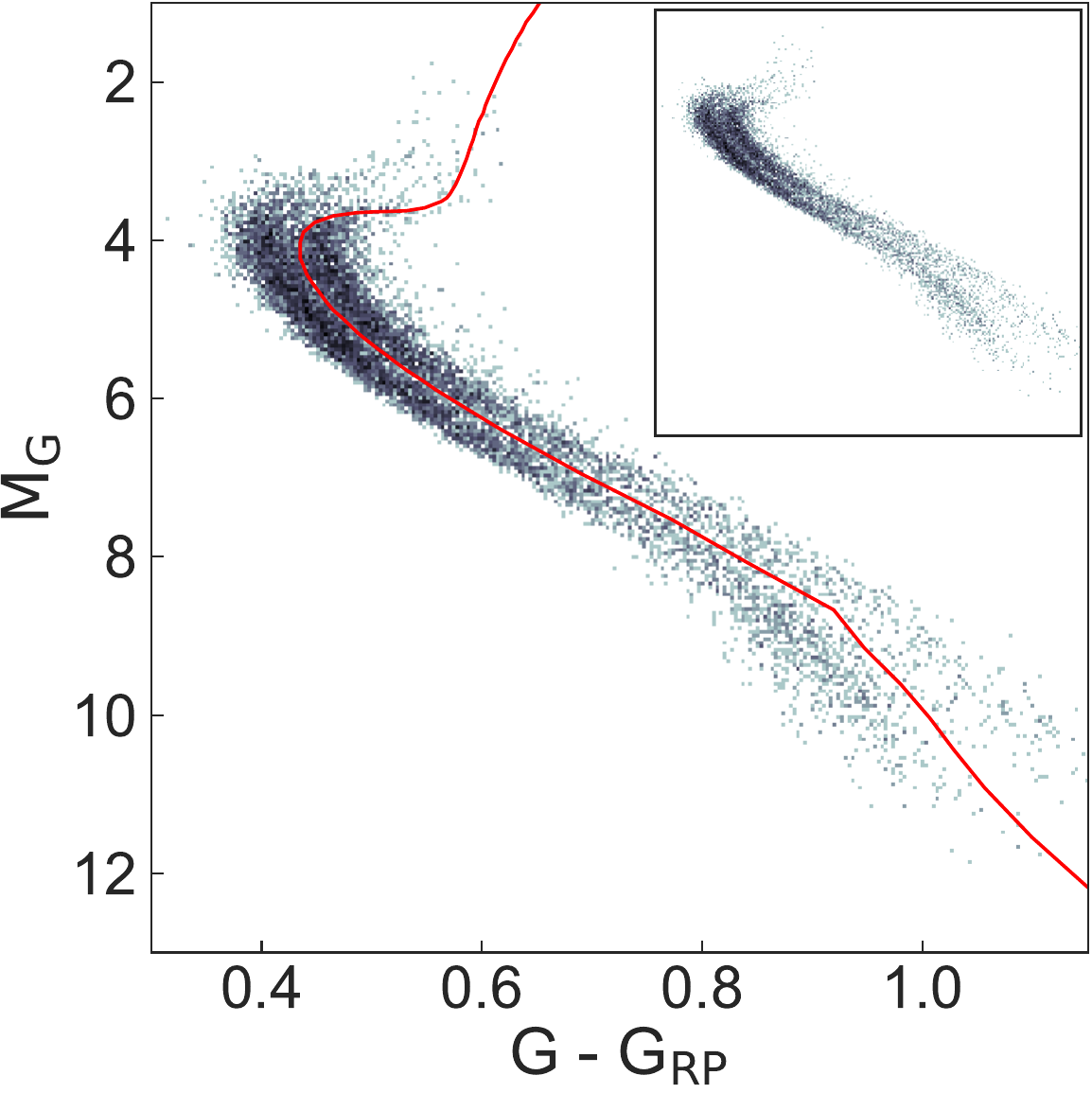}
    \caption{Hertzsprung-Russel diagram (HRD) of a local sample of stars with large tangential motions ($v_{\rm tan} > 200~{\rm km/s}$) and very high-quality parallaxes (${\tt parallax\_over\_error}>50$). Overlayed is an 11 Gyr age and [M/H]$=-0.5$ metallicity isochrone (red). This isochrone is shifted to the left by 0.01 mag in $G-G_{\rm RP}$ to split the two sequences that are shown. The inset shows the two sequences without the red line overplotted to make the gap between them appreciable.}
    \label{fig:HRD}
\end{figure}

\begin{figure}
    \centering
    \includegraphics[width=\hsize]{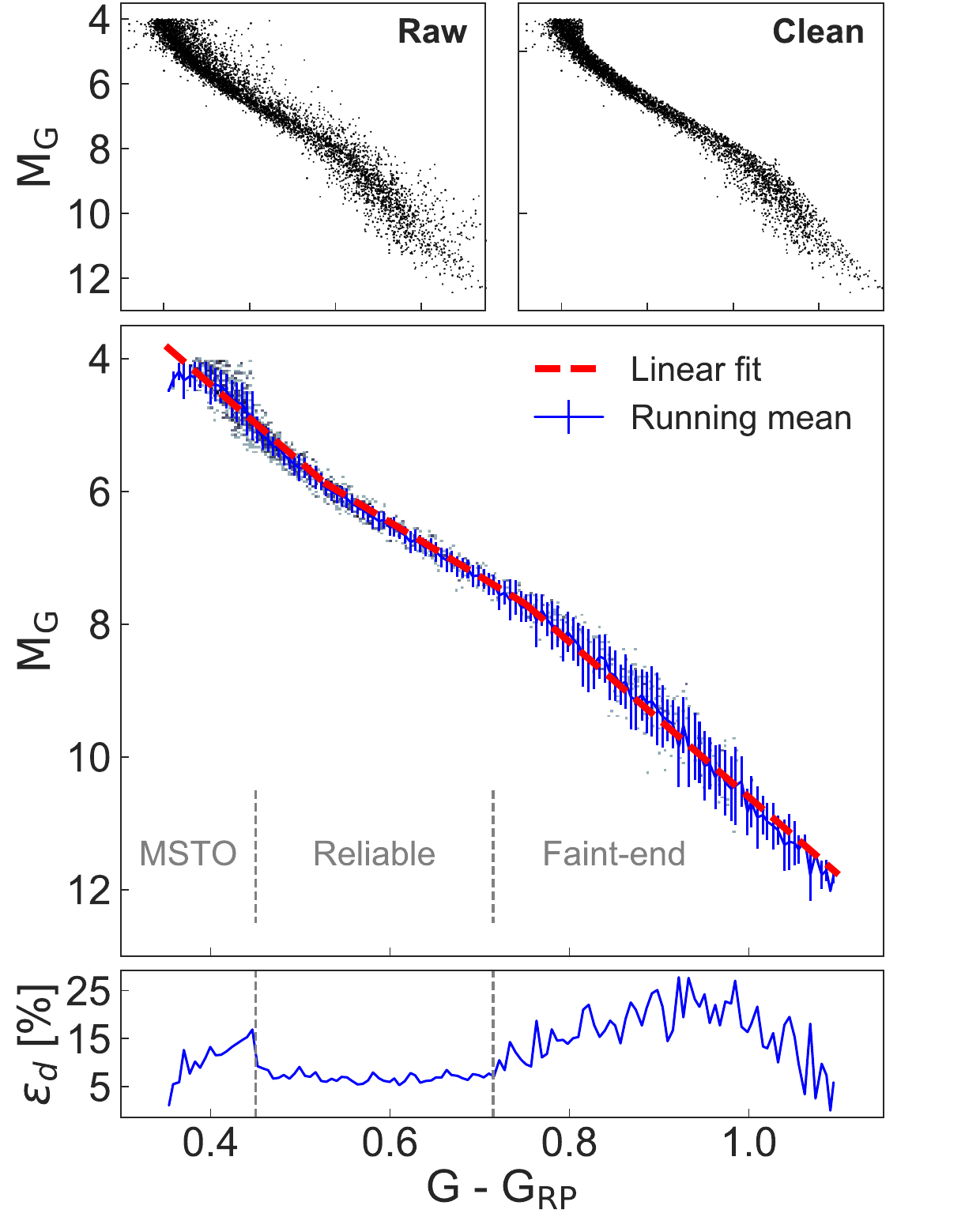}
    \caption{
    Top panels: raw data (left) and the cleaned data (right) that are used for the fit.
    Middle: two fits of the main sequence (MS).
    Bottom: error in distance is related to the measured thickness of the MS. The dashed lines indicate the range where the MS is reliable enough to calibrate photometric distances.}
    \label{fig:MSfit}
\end{figure}

Our method is contingent upon having a reliable fit for the absolute magnitude of halo MS stars as a function of colour. \cite{GaiaCollaboration2018Babusiaux} have shown that tangential velocities can be used to identify nearby halo stars (i.e. $v_\mathrm{tan} > 200~{\rm km/s}$). This set of high-$v_\mathrm{tan}$ stars is characterised by two sequences in the HRD, which are known as the blue and red sequence \citep[e.g.][]{GaiaCollaboration2018Babusiaux}.
The red sequence is kinematically reminiscent of the slower-rotating, hotter tail of the thick disc and the blue sequence of a classic halo that has a close to zero rotation \citep[e.g.][]{Koppelman2018, Haywood2018InDR2, DiMatteo2018, Gallart2019UncoveringGaia}. 

For the purpose of mapping the Galactic halo, we are mainly interested in the stars in the blue sequence. We expect the red sequence to be more important closer to the plane of the disc and in the inner Galaxy. It is currently not possible to avoid contamination from the red sequence (metallicity information could help, e.g. \citealt{ Gallart2019UncoveringGaia}). Its stars are, both kinematically and photometrically, too similar to the stars in the blue sequence. However, for the fitting, we strive to keep the contamination from the red sequence to a minimum. One way of doing this is to increase the cut in $v_{\rm tan}$ to a larger value. However, even at $v_{\rm tan}>300~{\rm km/s}$ the contribution of the red sequence is still $\sim 22\%$ \citep{Sahlholdt2019CharacteristicsDR2}.

Figure~\ref{fig:HRD} shows an HRD of all the stars in {\it Gaia DR2} that remain after imposing the quality cuts described in Sect.~\ref{sec:photqual} and two additional criteria: $({\tt parallax\_over\_error}>50)~\&~(v_{\rm tan}>200~{\rm km/s})$. For illustrative purposes we use here $v_{\rm tan}>200~{\rm km/s}$ rather than $300~{\rm km/s}$ because this brings out the two sequences better. However, for the fitting procedure we use $v_{\rm tan}>300~{\rm km/s}$. 

After the $v_{\rm tan}$ cut we impose a $M_G$-colour cut to remove the last bit of the red sequence contamination. For this cut, we use a synthetic isochrone that was first used by \cite{GaiaCollaboration2018Babusiaux} to describe the red sequence. The isochrone, overlayed in Fig.~\ref{fig:HRD}, describes a stellar population of a metallicity and age of ${\rm [M/H]}= -0.5$ and 11 Gyr. This isochrone is obtained from \cite{Marigo2017}, after enhancing the $\alpha$ elements by 0.23 \citep{Salaris1993}. The specific isochrone is chosen because it matches well with the red sequences as shown by \cite{GaiaCollaboration2018Babusiaux}, and it has the alpha-enhancement characteristic of the halo. We note that this isochrone is not fitted, but simply describes well the valley between the two sequences when shifted by 0.01 mag in $G - G_{\rm RP}$. All the stars to the right of the isochrone (i.e. those belonging to the red sequence) are removed. As a final quality cut we remove a handful of sources that are offset from (i.e. are below) the MS. For this cut we remove the stars with $((G - G_{\rm RP}<0.65)~\&~(M_G>8)) ~{\tt OR}~ ((G - G_{\rm RP}<0.8)~\&~(M_G>10))$.

For the final part of this section, we fit the cleaned MS in two ways: a simple three-component, piece-wise linear fit and a more accurate fit using a running mean and standard deviation of the MS.
The three-component fit has a straight forward parametrisation, which is ideal for the construction of the MS selection. We select the MS stars from the full dataset of {\it Gaia}, so a computationally efficient parametrisation is beneficial. On the other hand, the running mean is the most accurate fit, which is crucial for the calculation of the photometric distances. We have tested using a synthetic isochrone like the one overlaid in Fig.~\ref{fig:HRD} instead of fitting the MS. However, the isochrone does not perfectly trace the MS over the full colour range. The fit on the data is more precise, given that there are enough stars per bin.

The three components of the linear fit describe the MS in the absolute-magnitude ranges: $(4<M_G<6)$, $(6<M_G<8)$, and $(M_G>8)$. For the running mean we split the MS into 128 bins in the range of $0.35<(G - G_{\rm RP})<1.1$ and remove stars with $(M_G<4)$, which is roughly where the MSTO occurs. In each colour-bin, we calculate the mean absolute magnitude and the standard deviation. The resulting fit closely describes the width and amplitude of the absolute magnitude as a function of $G - G_{\rm RP}$. Both fits are run on a sample of high-$v_{\rm tan}$ stars with good parallaxes: $({\tt parallax\_over\_error}>50)~\&~(v_{\rm tan}>300~{\rm km/s})$.

Figure~\ref{fig:MSfit} shows both the three-component linear fit (red, dashed) and the running mean (blue). In the background, we show the sample that is fitted. The two fitting procedures agree well with the data. The top panels show the MS sample that was used in for the fitting before (left) and after (right) the photometric cleaning described in this section.


\subsection{Inferring distances for MS stars}\label{sec:dataphotdist}

We use the running-mean-fit from Sect.~\ref{sec:MSfit} to calibrate photometric distances. For each star we find the colour-bin in which it falls. The bin sizes are sufficiently small so we do not perform any interpolation. We assume that the absolute magnitude of the star is the same as the mean value found for the specific bin. Using Eqs.~\eqref{eq:dist} and \eqref{eq:edist}, we calculate the distance and its relative error.

The bottom panel of Fig.~\ref{fig:MSfit} shows the typical error, based on the width of the MS in absolute magnitude. Overall, the expected error in the photometric distance is quite small, averaging $7\%$ for a large fraction of the MS. The relative distance error of $10\%$ given for stars at the MS turn-off ($0.35 < {G-G_{RP} < 0.45}$) is somewhat misleading. Since the sequence here is close to vertical, the range in possible magnitudes is larger than what is indicated by the error bars. The fit of the MS aims to trace the faint part of the MSTO, since we do not fit for stars brighter than $M_G<4$. The fitted absolute magnitude is comparable to or smaller than the true absolute magnitude. Therefore, distances for MSTO stars that are far away from the fitted sequence are typically underestimated — and on average not overestimated. For comparison, if the intrinsic brightness of a source is underestimated by one magnitude, which is typical for the vertical extent of the MSTO, the distance will be underestimated by $37\%$. Another systematic bias that is not included in the relative error is the difference of half a magnitude that is typical for the offset between the red and blue sequence. This offset results in distances for red-sequence stars that are systematically underestimated by 20\%.

For faint MS stars, at ${G-G_{\rm RP} \approx 0.7}$, there is a break in the MS. This break is a known feature in the faint MS for low-mass stars \citep{Saumon1994CoolAtmospheres, Cassisi2000GalacticModels}. It is caused by a low effective temperature in combination with collisionally induced absorption. The feature is observed in globular clusters \citep[][who use it to determine the age of NGC 3201]{Bono2010OnCase} and in the Galactic bulge \citep{Zoccali2000The}. The width of the MS increases in the region redder than this break. As a result, the photometric distances are less reliable for stars with ${G-G_{\rm RP} > 0.715}$. The brightest of these $(M_G \sim 8)$ are only visible out to $\sim 4~{\rm kpc}$ (assuming a Gaia limiting magnitude of 21 in the G-band).
Our interest in the halo lies mainly with stars more distant than 4 kpc and with stars with reliable distances. Therefore, when using the newly calibrated photometric distances, we often constrain ourselves to stars in the range of $0.45 < {G-G_{\rm RP} < 0.715}$ only. This range is indicated in the bottom panel of Fig.~\ref{fig:MSfit} with vertical dashed lines. Of course, the distance calibration described in this section only works for MS stars. Therefore, the last step in the construction of our sample is to remove contamination from other types of stars (e.g. giants and white dwarfs).

\subsection{Selecting MS stars} \label{sec:rpm}

\begin{figure}
    \centering
    \includegraphics[width=\hsize]{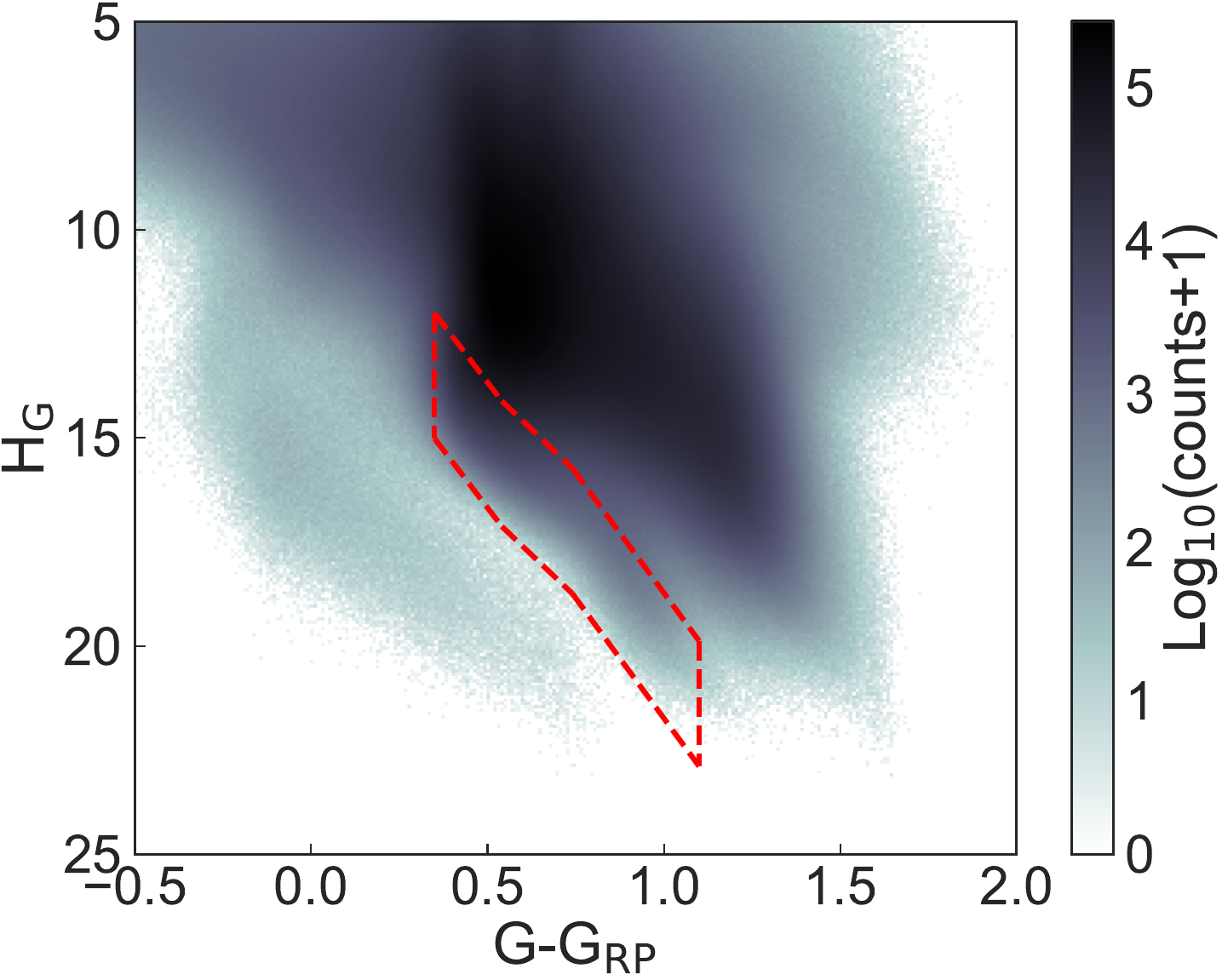}
    \caption{Reduced proper motion diagram for all sources in {\it Gaia} that pass the quality cuts of Sect.~\ref{sec:photqual}. The sample of sources within the red lines are selected to be tentatively halo stars. The red lines are drawn based on the three-component fit of the MS. The location of the box is chosen to select halo stars with $ 200~{\rm km/s}< v_{\rm tan} < 800~{\rm km/s}$.}
    \label{fig:RPMselection}
\end{figure}

The method of selecting halo MS stars in the RPM diagram is best understood when visualised. Figure \ref{fig:RPMselection} shows the RPM as a function of colour for the full {\it Gaia} data set after imposing the quality cuts from Sect.~\ref{sec:photqual}. To find the location of MS stars with a large tangential velocity we place the three-component linear fit from Sect.~\ref{sec:MSfit} on top of the density map in Fig.~\ref{fig:RPMselection}. We add the $v_{\rm tan}$-based offset from Eq.~\eqref{eq:Hg2} to the line. The upper line of the box in Fig.~\ref{fig:RPMselection} is given by the fit plus an offset of $200$ km/s, the lower line has an offset of $800$ km/s. Therefore, the stars that fall between the horizontal lines are those that, based on their location in the RPM diagram, are on the MS and have a velocity in the range of $200~{\rm km/s} < v_\mathrm{tan}<800~{\rm km/s}$. The vertical lines of the box are set to $0.35 < {G - G_{\rm RP}} < 1.1$. The blue limit $(0.35 < {G - G_{\rm RP}})$ is chosen because there are no MS stars bluer than this in the halo, see for example Fig.~\ref{fig:HRD}. At the other end, we truncate the selection at ${G - G_{\rm RP}} = 1.1$ because this is approximately where the MS ends. We note that the truncation is beyond the red limit where the distances become less reliable. However, these stars are still likely halo stars and therefore we add them to the sample.


\subsection{Final quality checks}

\subsubsection{Removing white dwarfs}\label{sec:WDs}
\begin{figure}
    \centering
    \includegraphics[width=\hsize]{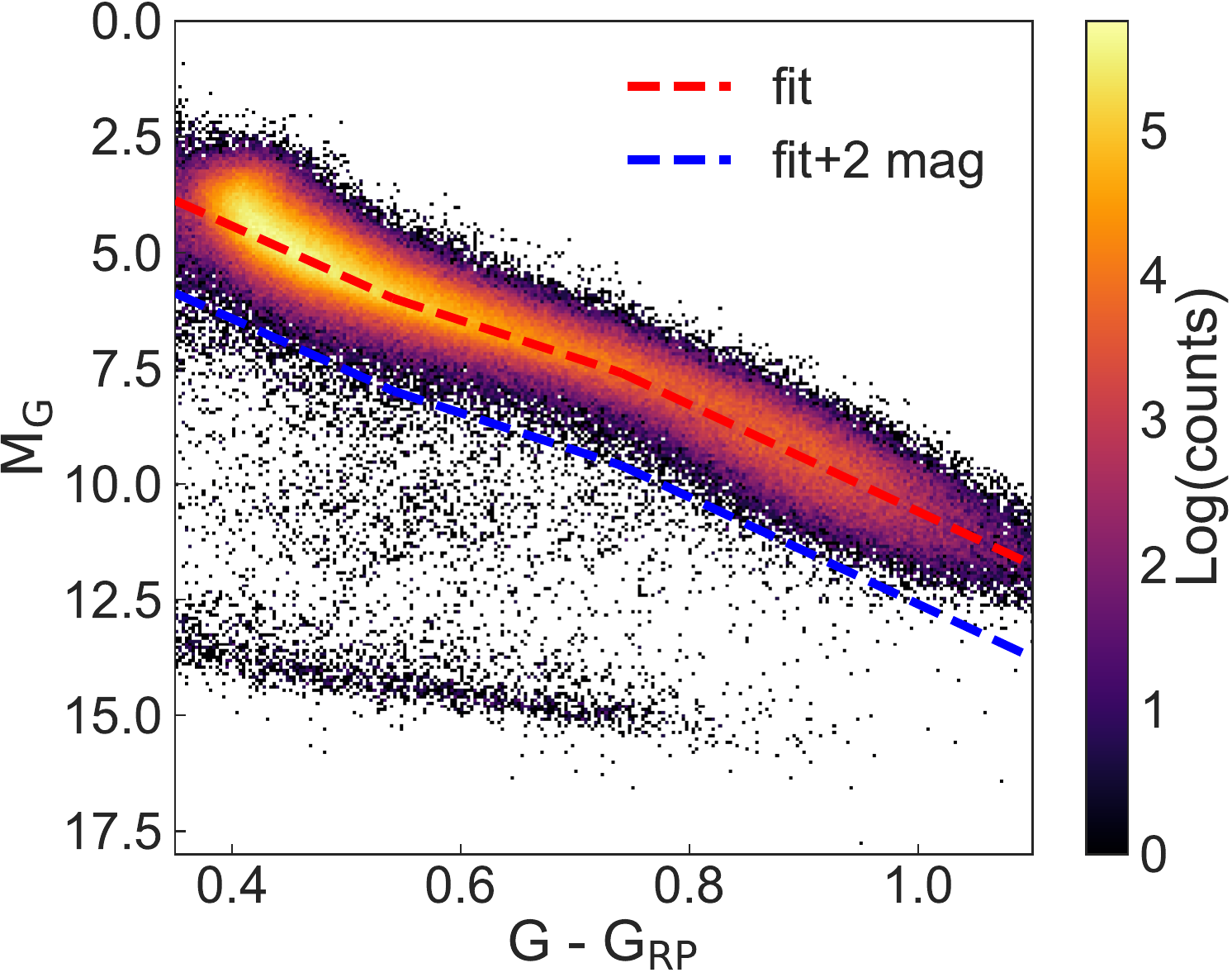}
    \caption{HRD of sources in the MS selection with precise parallaxes (${\tt parallax\_over\_error} > 5$). The three-component fit of the MS of Sect.~\ref{sec:MSfit} is shown as a red, dashed line and with an offset of 2 mag as blue line. Contamination of white dwarfs and other faint sources is reduced by removing all the stars below the blue line.}
    \label{fig:WDremoval}
\end{figure}

Figure \ref{fig:WDremoval} shows the HRD of a subset of the RPM sample with ${\tt parallax\_over\_error}>5$, where the $M_G$ is calculated using the parallaxes. The red line is the three-component fit that is also shown in Fig.~\ref{fig:MSfit}, and the blue is this same line, but offset by 2 mag in $M_G$. Below the blue line, there is contamination from faint sources. Amongst these sources is a population of white dwarfs, visible at $12.5<M_G<15$. The purpose of the blue line is to filter this contamination, that is, we remove all the sources below the line (and have ${\tt parallax\_over\_error}>5$). 

Based on the number of sources with reliable ${\tt parallax\_over\_error}>5$, we estimate that the contamination is at most $0.2\%$. White dwarfs are intrinsically faint objects, the brightest few in our sample have an absolute magnitude of $M_G \sim 13$. As a result, we expect no contamination of white dwarfs farther out than $\sim 0.40$ kpc, beyond which they fall below the magnitude limit of {\it Gaia}. Because white dwarfs can only be observed in the close vicinity of the Sun, we expect that all of them have relatively good parallaxes. Therefore our filter shown in Fig.~\ref{fig:WDremoval} should remove all contamination from white dwarfs. 

Giants are an unlikely source of contamination because of their intrinsic brightness and even the uncertainties in the proper motions or photometry are not large enough. Therefore, we estimate that the fraction of contamination of non-MS stars in our final RPM sample is negligible.

\subsubsection{Quality of photometric distances} \label{sec:qualphotdist}
\begin{figure}
    \centering
    \includegraphics[width=\hsize]{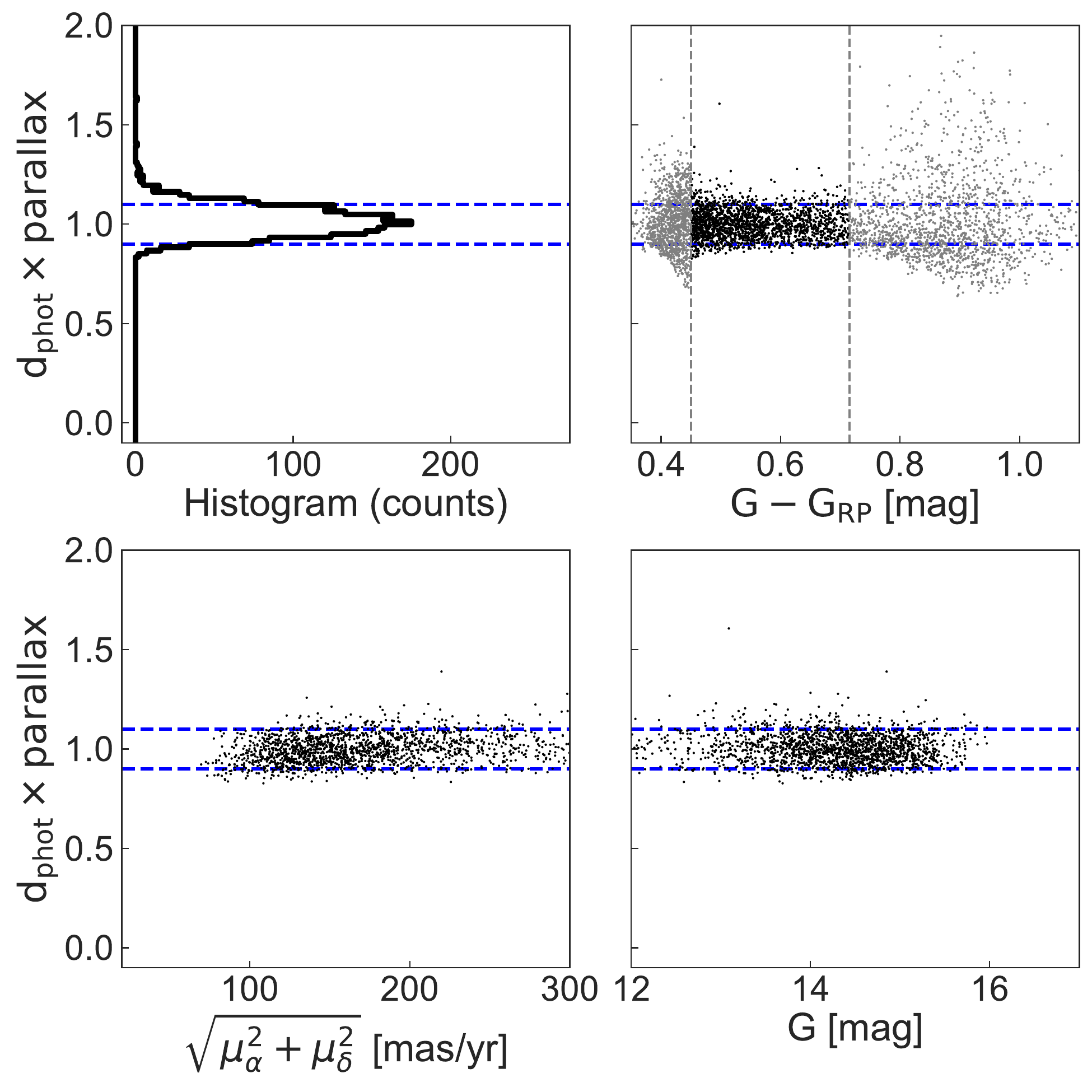}
    \caption{
    Inspection of the quality of the photometric distances compared to the parallax (top, left), and as a function colour (top, right), proper motion (bottom, left), and extinction corrected $G$-magnitude (bottom, right). As a reference, we overlay dashed lines indicating a difference of $\pm 10\%$ in the distance. The quality of the photometric distances does not depend on any of the values showed here, as is to be expected.}
    \label{fig:photdistquality}
\end{figure}

We test the quality of the photometric distances in Fig.~\ref{fig:photdistquality}, which shows photometric distance quality for the same sample of stars used for fitting the MS. The quality is displayed as a function of the colour ${G-G_{\rm RP}}$, the amplitude of the proper motion, and ${G}$-magnitude. The vertical dashed lines in the top right panel of Fig.~\ref{fig:photdistquality} indicate the limit for reliable distances that is described in Sect.~\ref{sec:dataphotdist}. The stars that are outside of this range (in grey) are not included in the other panels. The typical parallax-distance error in the sample that is shown can be neglected, it is $\lesssim 2\%$. These figures confirm that there is no dependence and that the distance derivation works properly. The photometric distances agree well with the parallaxes. In each panel of Fig.~\ref{fig:photdistquality}, we overlay two blue horizontal dashed-lines that correspond to photometric distances that are 10\% off from the parallaxes.

\subsubsection{Cross-matches with spectroscopic surveys}

\begin{figure}
    \centering
    \includegraphics[width=\hsize]{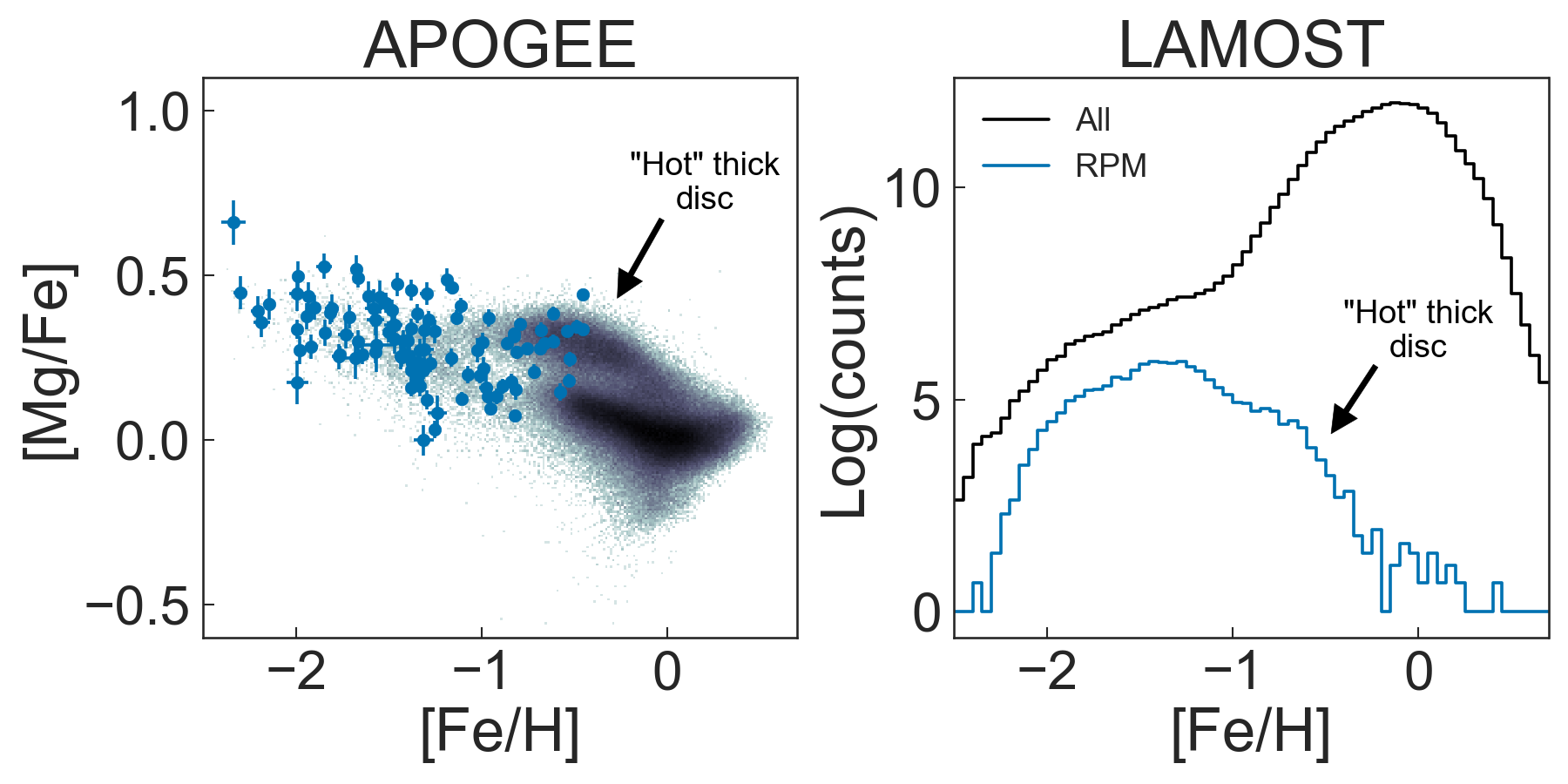}
    \caption{Chemical abundances (left) and metallicities (right) of stars included in the RPM sample as derived from cross-matching to APOGEE DR16 and to LAMOST DR4. In both panels, we denote the distribution of the full samples provided by each of these surveys, after quality cuts, in black, while in blue we denote the subset that overlaps with the RPM sample. The vast majority of these stars are clearly all metal-poor, as expected for halo stars.}
    \label{fig:specAPLA}
\end{figure}

As a last check before investigating the properties of the RPM halo sample, we briefly explore the chemical properties of stars cross-matched to existing spectroscopic surveys. Figure~\ref{fig:specAPLA} shows the results of the overlap of the RPM sample with APOGEE DR16 \citep{Ahumada2020} and with LAMOST DR4 \citep{Cui2012}. 

For the comparison with APOGEE we plot in the left panel of Figure~\ref{fig:specAPLA}  the characteristic [Mg/Fe] vs [Fe/H] diagram, for the full dataset in black and with the halo RPM stars in blue. We show here only stars that pass the following quality criteria: ${\tt VERR} < 0.2$, ${\tt SNR} > 70$, ${\tt TEFF} > 3700$, {\tt STARFLAG} bitmask values of $0$, $3$, or $4$ and {\tt ASPCAPFLAG} bitmask values of $10$ or $23$ where {\tt SNR} is the catalogue parameter for the signal-to-noise ratio \citep[these criteria are based on][]{Hayes2020}. This figure shows that stars overlapping the RPM sample are all metal-poor and mostly lie on the halo sequence. Only a handful of stars lie on the high-$\alpha$ disc sequence. These stars are most likely part of the `hot' thick disc (i.e. the `red sequence' discussed in Sect.~\ref{sec:dataphotdist}). These stars have halo kinematics and are unavoidable in a kinematically selected halo sample such as the RPM sample explored here. 

The panel on the right in Fig.~\ref{fig:specAPLA} shows the metallicity distribution for LAMOST in black, and for the overlap with the RPM sample in blue. In this case we have imposed the following quality cuts: ${\tt snri}>20$ and ${\tt snrg > 20}$. Again, the RPM subset is clearly more metal-poor than the full sample. The overlapping subset (blue curve) shows a secondary bump at [Fe/H] $\sim -0.8$ that most likely corresponds to stars on the `hot' thick disc sequence that are also visible in the cross-match with APOGEE. This interpretation is supported by the fact that when using the LAMOST line-of-sight velocity measurements, stars with [Fe/H] $> -0.9$ have more disc-like kinematics than the more metal-poor stars (which all have halo-like motions). This result is fully consistent with the analyses of the halo and `hot' thick disc stars of, for example, \cite{Haywood2018InDR2} and \cite{DiMatteo2018}. We have also performed similar tests with data from GALAH DR2 \citep{Buder2018}, RAVE DR5 \citep{Kunder2017}, and APOGEE-astroNN \citep{Leung2018} and reached similar conclusions as those presented here.

\begin{figure}
    \centering
    \includegraphics[width=\hsize]{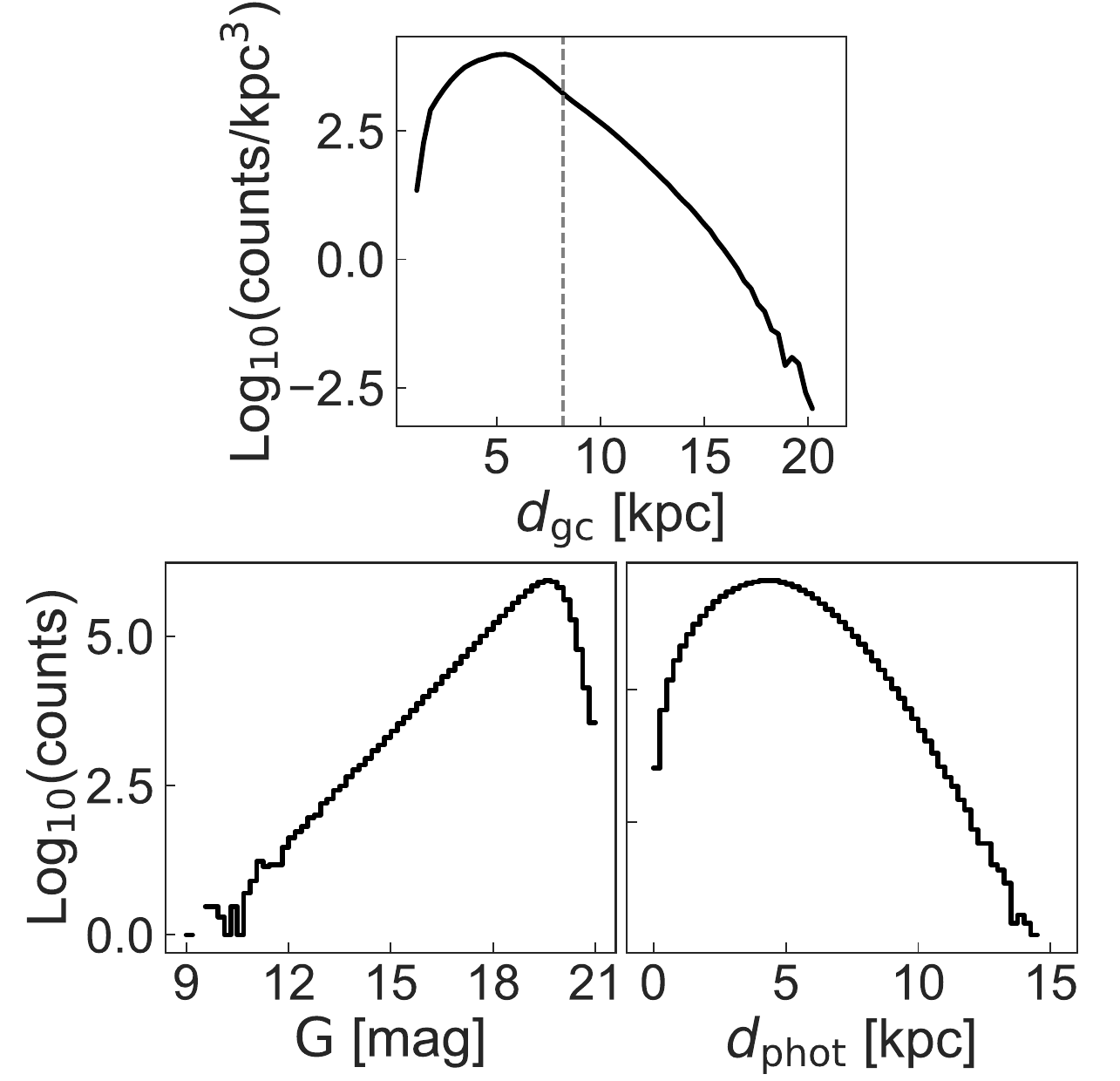}
    \caption{Top: Distribution of galactocentric distances. The dashed line indicates the location of the Sun. Bottom: Distribution of G-magnitudes (left) and photometric distances (right) of the stars in the RPM sample.}
    \label{fig:hists}
\end{figure}
\begin{figure*}[!ht]
    \centering
    \includegraphics[width=\hsize]{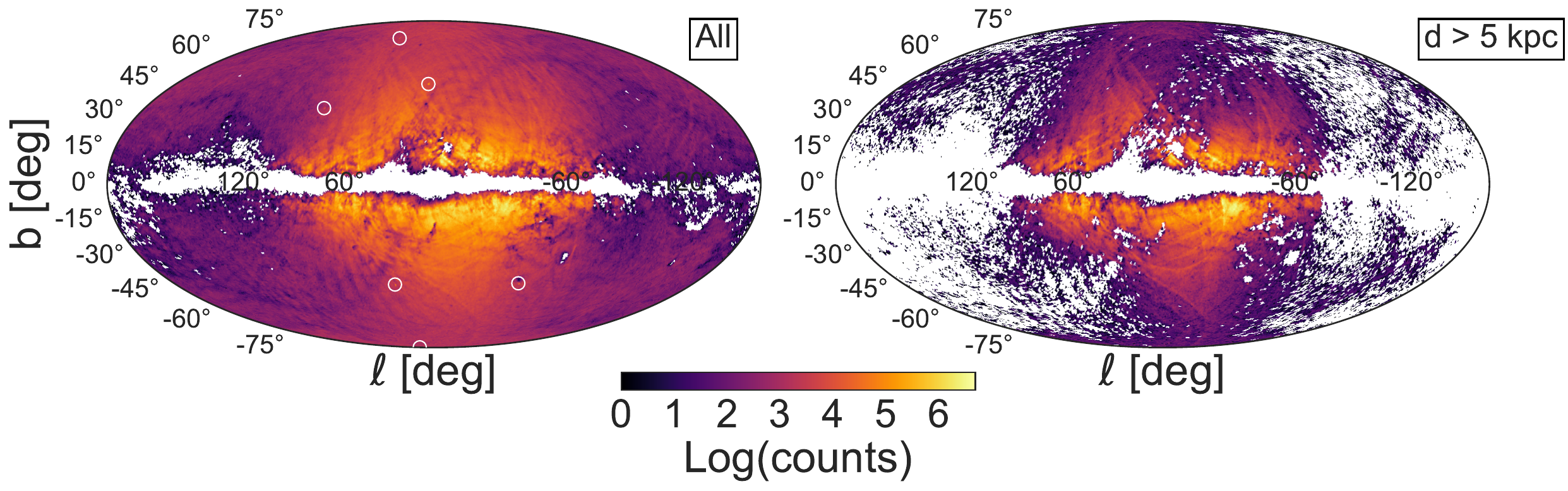}
    \caption{Sky-maps of RPM selected sample of halo stars. The left panel shows the full RPM sample, and the right panel shows a selection of the 35\% most distant stars. Strong signatures of the {\it Gaia} scanning pattern are visible in both panels. Sources in the plane of the disc are filtered by our quality cuts. The white circles mark the location of six globular clusters that are picked up by the RPM method and which are clearly visible when zooming in.}
    \label{fig:radecmap}
\end{figure*}
\begin{figure*}
    \centering
    \includegraphics[width=\hsize]{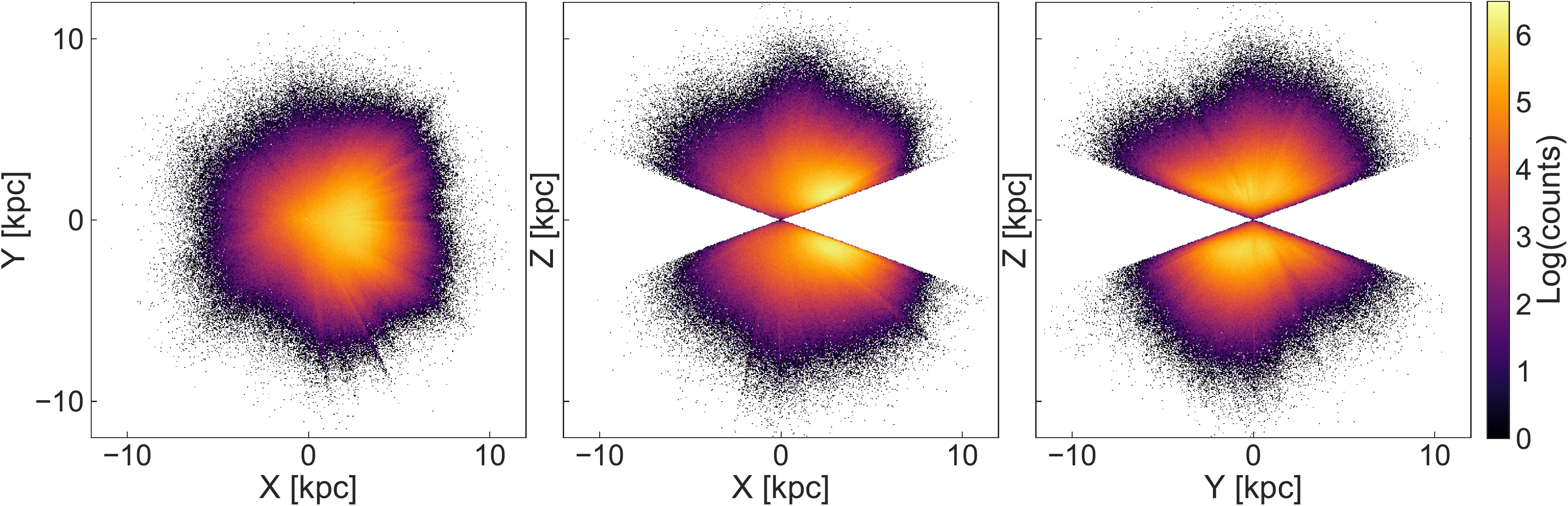}
    \caption{Spatial distribution of the RPM sample in heliocentric Cartesian coordinates. The sample is distributed uniformly, and the density of sources decreases with heliocentric distance, as is expected for a magnitude-limited sample. A combined effect of dust extinction and the scanning pattern of {\it Gaia} creates stripes. Sources with $|b|<20^\circ$ have been removed (i.e. highly reddened stars located close to the disc).}
    \label{fig:xyzmap}
\end{figure*}

\section{Spatial distribution of the RPM sample}\label{sec:spatdist}

The final sample obtained using the above described procedures comprises $11~711~399$ tentative MS halo stars. The subset with reliable photometric distances (i.e. stars in the range $0.45 < {G - G_{\rm RP}} < 0.715$, which excludes MSTO and faint-end stars) comprises $7~117~555$ stars.

We now inspect the spatial distribution of the stars in the RPM sample and check for signs of substructure. Because MS stars are intrinsically faint objects, {\it Gaia} can only observe them up to $\sim 16$ kpc, assuming the brightest star has ${\rm M_G}\sim5$ and {\it Gaia}'s limiting magnitude in ${G}$ is $\sim21~{\rm mag}$. However, at the faint-end, {\it Gaia} is far from complete. Only $7384$ sources $(\sim 0.1\%)$ in the sample have a photometric distance larger than 10~kpc - the median photometric distance of the sample is $4.39$~kpc.

Figure~\ref{fig:hists} shows the distribution of the galactocentric distance, G-magnitude, and photometric distance. The top panel shows a centrally concentrated distribution. Close to $83\%$ of the stars are located inside of the solar radius $(d_{\rm gc} < 8.2~{\rm kpc})$. This high concentration towards the centre is to be expected, the stellar halo is known to have a steep density profile \citep[e.g.][]{Juric2008THEDISTRIBUTION, Deason2011TheSmooth}.

Figure \ref{fig:radecmap} shows the distribution of the stars in the RPM sample in a Mollweide projection, colour-coded by the logarithm of the number of stars per pixel. The maps are created at a healpix level of 7, resulting in $196~608$ equal area pixels.
The left panel shows the distribution for the complete sample and the right panel shows all sources with a photometric distance larger than 5 kpc ($\sim 35\%$ of the sample). Several globular clusters are visible in both panels as small yellow dots, they are highlighted with white markers. These globular clusters are all nearby, they are NGC 7099, NGC 362, NGC 5904, NGC 6341, NGC 5466, and NGC 288.

{\it Gaia}'s scanning-pattern \citep[c.f.][where these patterns are shown and discussed]{Lindegren2018, Arenou2018} is prominent in both maps but is most clearly seen in the right panel. The sinusoidal band with an amplitude of $\sim 60^\circ$ in $b$ is a known artefact in the {\it Gaia} DR2 data related to insufficient subtraction of zodiacal light \citep[c.f. Fig.~18 of][]{ Evans2018GaiaValidation}. Besides these systematic effects, the right panel shows significant signs of incompleteness, most clearly apparent in the number of bins with zero sources (white pixels). The pixels with no stars, in the left panel, are mostly coinciding with high-extinction regions. See, for example, the disc region, but also the area near the Magellanic Clouds.

\begin{figure*}
    \centering
    \includegraphics[width=\hsize]{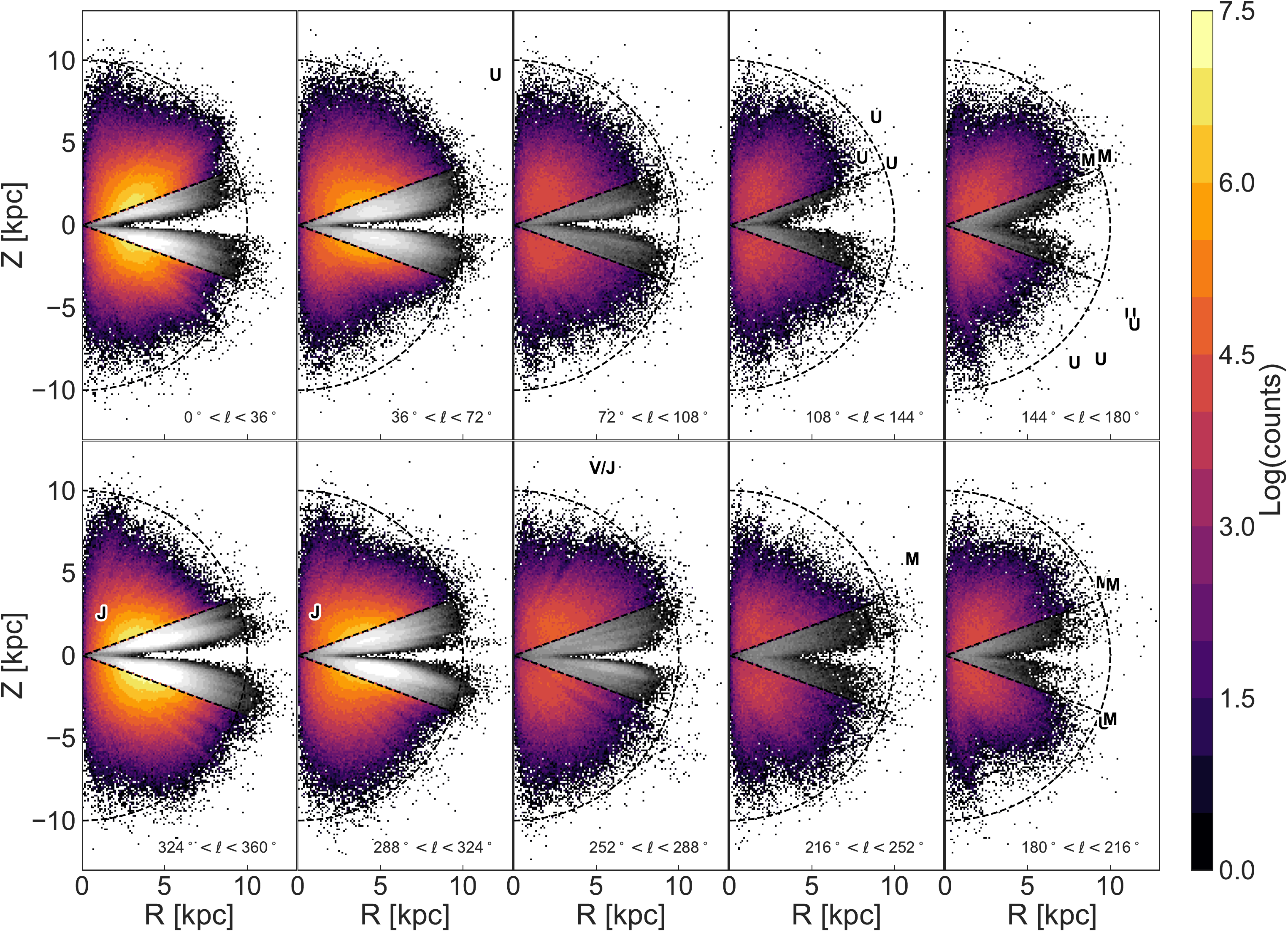}
    \caption{Spatial distribution in wedges of $\Delta\ell = 36^\circ$, projected in cylindrical heliocentric $(R,Z)$ coordinates. Low latitude $(|b|<20^\circ)$ areas are desaturated because these are likely contaminated by thick disc stars. The labels correspond to the overdensities found by \cite[][i.e. their Table 4]{DeJong2010MAPPINGPHOTOMETRY}: Virgo (V), Monoceros (M), (6.5, 1.5) (J; detected originally by \cite{Juric2008THEDISTRIBUTION}), and the unlabelled structures (U). These overdensities lie mostly just beyond our sample's reach, but might become visible with a similar analysis with upcoming {\it Gaia} data releases.}
    \label{fig:longitude-slices}
\end{figure*}

\begin{figure}
    \centering
    \includegraphics[width=\hsize]{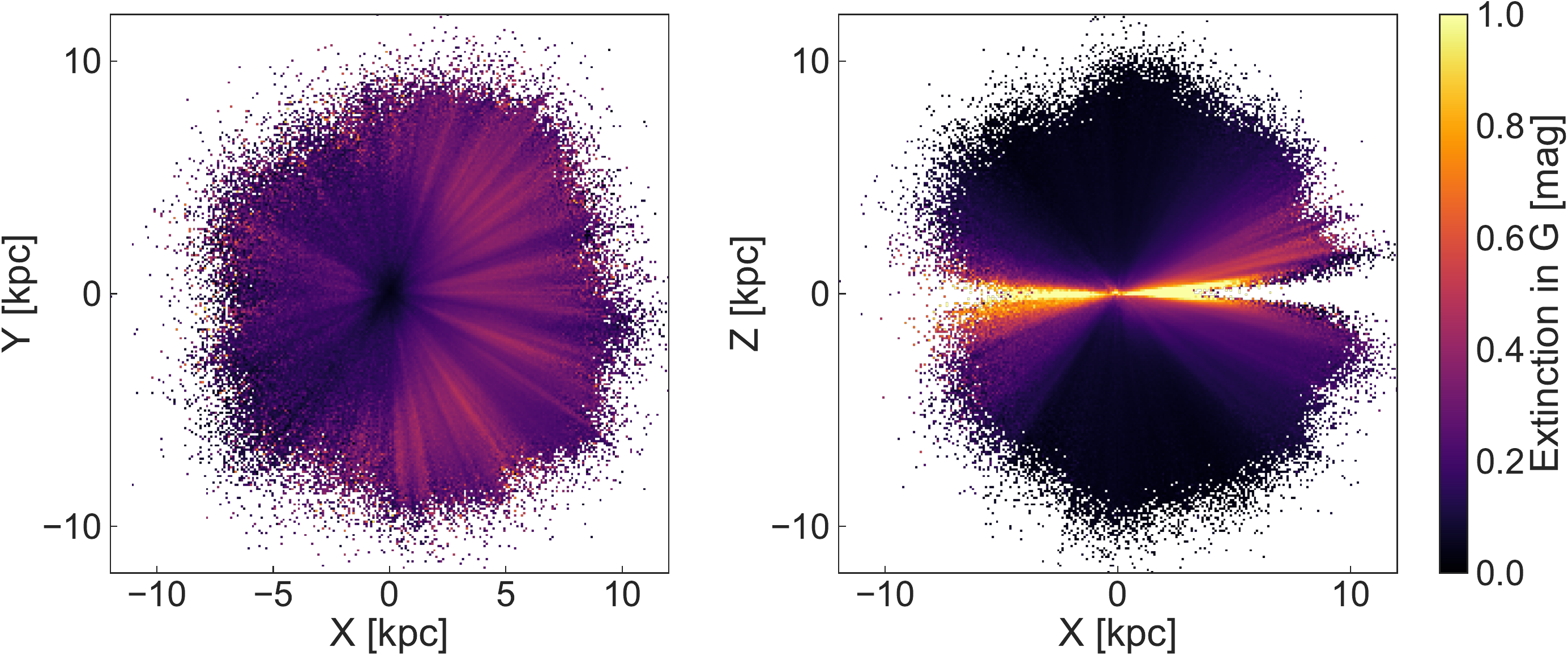}
    \caption{Spatial distribution of the RPM sample projected onto the $XY$-plane (left) and $XZ$-plane (right). The sample is binned in $256\times256$ bins, each bin is colour-coded by the mean extinction in $G$-mag. As expected, low-latitude regions are the most affected. The extinction creates interesting features as a function of galactic $\ell$ (i.e. the fingers of God features).}
    \label{fig:xyz-Ext}
\end{figure}

The spatial distribution of the sample, shown in Fig.~\ref{fig:xyzmap}, displays a similar level of structure. The coordinate system is oriented such that the Galactic Centre is in the positive $X$-direction and the rotation of the disc is in the positive $Y$-direction. To mitigate the contamination from highly extinct sources and from the thick disc we have removed all the stars with $|b|<20^\circ$. The combined effects of the remaining dust extinction, the scanning pattern of {\it Gaia}, and the errors in the photometric distance create the radial features seen in the figure.

Maps such as those shown in Fig.~\ref{fig:xyzmap} project 3D-structure onto a 2D-plane. As a result, most small-scale structure is smoothed out. A simple method to inspect the internal 3D-structure is to minimise the smoothing effect by dividing the sample in thin slices. In Fig.~\ref{fig:longitude-slices} we project wedges in $\ell$ in heliocentric cylindrical $(R,Z)$ coordinates, each wedge is $36^\circ$ in width. The maps are binned $(128\times256~{\rm bins})$ and coloured by the logarithm of the number of stars per bin. Low-latitude areas $(|b|<20^\circ)$ are highly affected by extinction, which introduces systematic errors in the distance estimate. Therefore, these areas are coloured in greyscale to focus the attention to the less affected parts. We overlay overdensities found in SDSS \cite[i.e. Table~4 of][]{DeJong2010MAPPINGPHOTOMETRY}, see the caption of the figure for more information. Most of the large overdensities such as the Virgo Over Density (VOD) \citep{Newberg2002THEWAY, Juric2008THEDISTRIBUTION, Bonaca2012UPDATEOVERDENSITY}, the Hercules-Aquila Clouds (HAC) \citep{Belokurov2007THECLOUD, Simion2014StrongCloud}, TriAnd \citep{Majewski2004DetectionRegion, Rocha-Pinto2004, Deason2014TriAndHalo} and similar structures \citep[e.g.][]{DeJong2010MAPPINGPHOTOMETRY, Grillmair2016StellarHalo} are too distant to be detected in our sample.

\begin{figure*}
    \centering
    \includegraphics[width=\hsize]{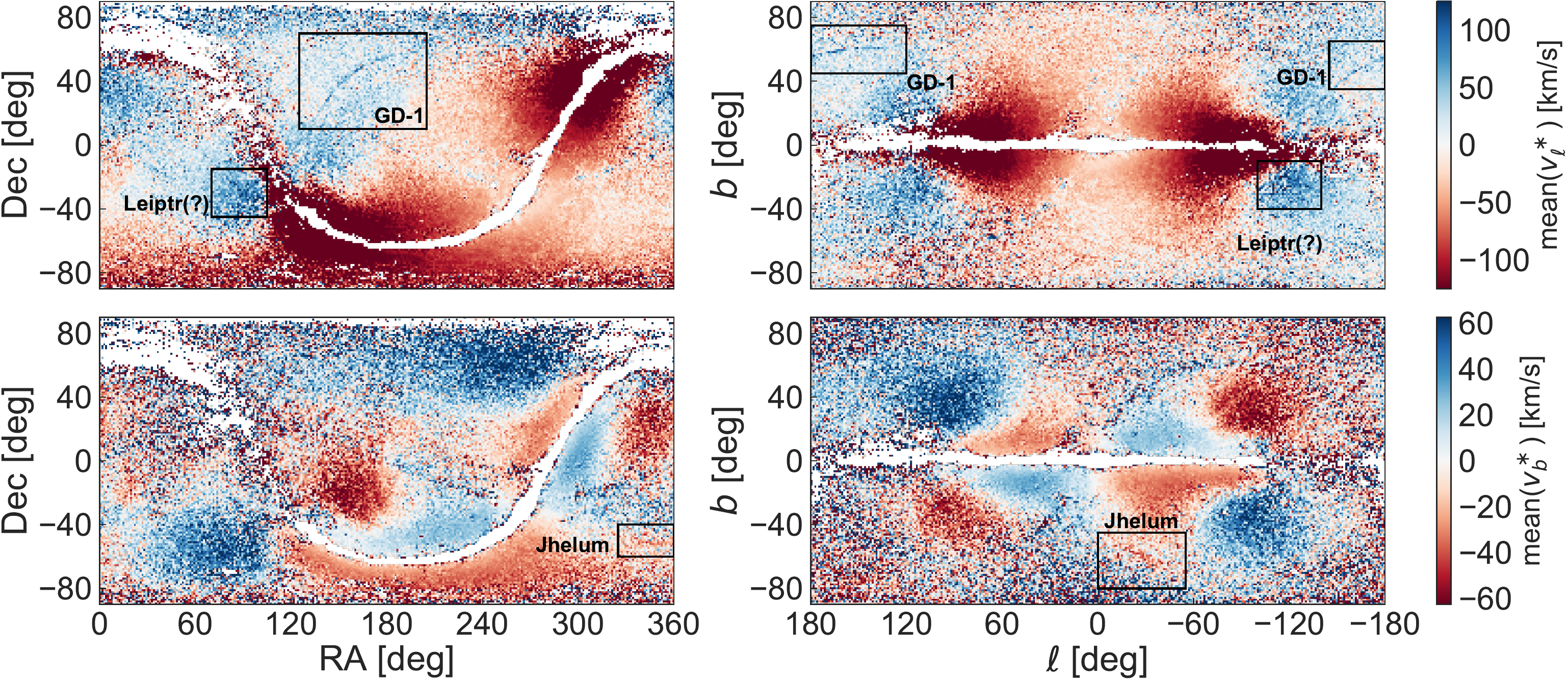}
    \caption{Distribution of the mean, solar-motion-corrected $v_\ell$ (top) and $v_b$ (bottom) velocities binned on the sky for distant halo stars $(d>6~{\rm kpc})$. Large-scale streaming motions as well as several small streams stand out. This map includes MSTO stars, which might have large uncertainties in the distances. See Appendix~\ref{sec:Apvelmaps} and Fig.~\ref{fig:old-vlvb} for a version without MSTO stars.}
    \label{fig:vlvb}
\end{figure*}

The distribution of stars is not isotropic as, for example, is shown in the lower-left panel of Fig.~\ref{fig:longitude-slices} which depicts a strong north-south asymmetry. This asymmetry is likely caused by incompleteness of {\it Gaia} because of its scanning pattern. There is an asymmetry in the number of faint stars ($G >19$) when comparing the source number counts in that specific wedge for $b>0$ versus the $b<0$. Some of the asymmetries are related to the photometric quality cuts that we impose, or are linked to the {\it Gaia} scanning pattern. They, for example, correlate with the {\it Gaia} DR2 catalogue parameter {\tt visibility\_periods\_used}. The complexity of the structure in the data makes it non-trivial to compare the structure against a smooth model, which makes the significance of the structure present in the maps here unclear.

We check for spatial trends with extinction in $G$-mag in Fig.~\ref{fig:xyz-Ext} where the mean extinction is shown in bins projected on the $XY$- and $XZ$-plane. As expected, the extinction strongly correlates with Galactic latitude $b$. Stars with low-latitude (the disc region) are strongly affected by extinction. The pattern in the $XY$-plane is less intuitive to understand. The low-latitude features in $b$ correlate non-uniformly with features in $\ell$. Towards the anti-centre $(X<0)$ the extinction is less significant. From the right panel, it is clear that a significant fraction of the high-extinction sources is easily removed by applying a simple cut in latitude $b$. Although these low-latitude stars are removed in Fig.~\ref{fig:xyzmap}, some of the stripes still seem to correspond to the high-extinction features.

\section{Velocity content of the RPM sample}\label{sec:veldist}

The spatial distribution of the RPM sample shows many features embedded in what is likely a smooth background. Much of the structure that is visible could be related to incompleteness and selection effects. In the next part we combine the spatial information with that encoded in the proper motions to filter halo structures from the smooth background.
This will be easier when such structures move sufficiently different
from the `background' (in the particular region of the sky). The degree of distinction will depend also on the magnitude of the velocity errors.

\subsection{Binned velocity moments}\label{sec:velmom}
A powerful, yet simple, tool is to bin moments of the velocity distribution on the sky.
Using the photometric distances, we convert the proper motions to tangential velocities using
\begin{equation}\label{eq:vlvb}
    v_{j} = 4.74057~{\rm km/s}~
    \bigg(\frac{\mu_{j}}{\rm mas/yr}\bigg)~
    \bigg(\frac{d}{\rm kpc}\bigg),
\end{equation}
where $j=(\ell,b)$. These tangential velocities can be corrected for the solar reflex motion
\begin{equation}
    v_{j}^\ast = v_{j} + v_{j,\odot},
\end{equation}\label{eq:vlbsolcor}
using
\begin{subequations}
\begin{equation}
    v_{\ell,\odot} = 
    -U_\odot \sin{\ell} + 
    (V_\odot+v_{\rm LSR})\cos{\ell},
\end{equation}
\begin{equation}
    v_{b,\odot} = 
    W_\odot\cos{b} - 
    \sin{b}\cdot(U_\odot\cos{\ell} + 
    (V_\odot+v_{\rm LSR})\sin{\ell}),
\end{equation}\label{eq:vlvbsun}
\end{subequations}
where we use the \cite{Schonrich2010} solar motion $(U_\odot,V_\odot,W_\odot) = (11.1,12.24,7.25)~{\rm km/s}$ and the \cite{Mcmillan2017} motion of the local standard of rest (LSR) $v_{\rm LSR} = 232.8~{\rm km/s}$. The resulting tangential velocities are in the Galactic rest frame, but as observed from the solar position.

For our sample of MS stars, the median velocity errors taking into
account the error in the distance as well as that in the proper
motions, are $\epsilon(v_\ell) \sim 22$~km/s and
$\epsilon(v_b) \sim 15$~km/s. For MSTO stars and those farther away
than 6~kpc, the median errors are larger, namely
$\epsilon(v_\ell) \sim 42$~km/s and $\epsilon(v_b) \sim 28$~km/s.

Figure~\ref{fig:vlvb} shows the mean $v_\ell^\ast$ and $v_b^\ast$ velocities binned on the sky. We only consider stars with a photometric distance $d>6~{\rm kpc}$ because we are interested in picking up distant streams (nearby streams will not appear as coherent, thin structures on the sky). Besides large-scale velocity patterns, which we discuss below, a few stream-like features stand out because members of a stream move in the same direction and with similar mean velocity, or at least sufficiently distinct from that of the background given the velocity errors. The most conspicuous structure is the GD-1 stream \citep{Grillmair2006DETECTIONSURVEY}, but also Jhelum is apparent as well as tentatively Leiptr \citep[c.f. Fig.~6 of][]{Ibata2019TheGalaxy}.

As mentioned above, we have included all of the stars in our sample also those near the MSTO, even though their distances will on average be underestimated by up to $\sim40\%$ (see Sect.~\ref{sec:qualphotdist}). Yet, because they are intrinsically brighter they allow us to probe the farthest into the halo. This means that the mean velocities shown in the figure might not be very accurate, yet including these MSTO stars allows us to probe distant streams whose motions are sufficiently different from those of the background stars in a similar portion of the sky, as appears to be the case for GD-1 and the Jhelum streams.

The large all-sky patterns that are seen in Fig.~\ref{fig:vlvb} are reminiscent of a rotation signal (particularly in $v_\ell^\ast$), and this is plausibly related to contamination from the hot thick disc. The pattern in $v_\ell^\ast$ aligns perfectly with the signal of the solar motion around the Galactic Centre (i.e. the change of sign at $\ell = \pm 90^\circ$), further supporting the interpretation of rotation.

Although the correction for the solar motion, and in particular the contribution from the LSR motion, could potentially affect these velocities (since the correction is dependent on the photometric distance), the effect is unlikely to be significant. This view is supported by the large-scale pattern seen in $v_b^\ast$ in the figure, which cannot be due to the solar motion correction (because of its dependence on galactic longitude).

\begin{figure}
    \centering
    \includegraphics[width=\hsize]{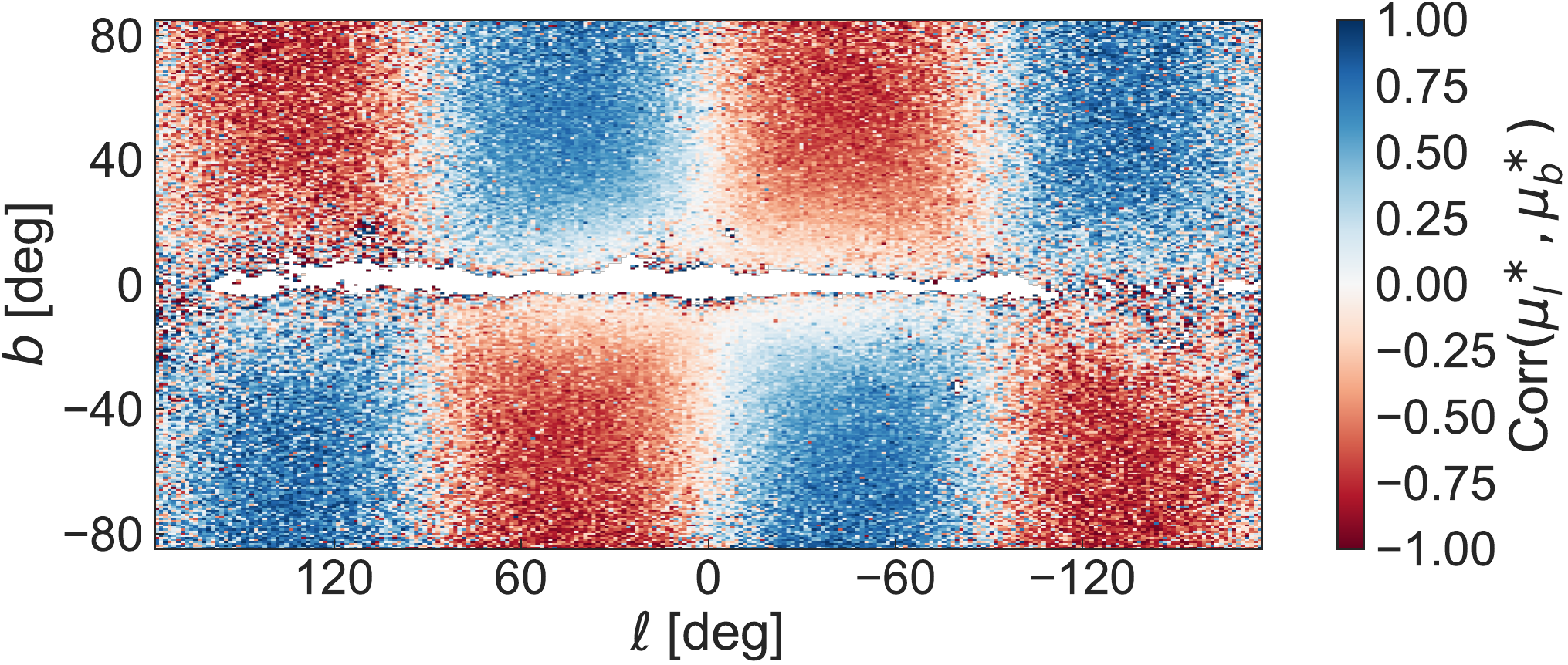}
    \caption{Mean Pearson correlation coefficient of the proper motions projected on the sky after correction of the solar reflex motion. The signal that is present was first observed in a sample of RR-Lyrae stars \citep[][see their Fig.~5]{Iorio2019TheMerger}. A radially anisotropic halo explains the pattern, mainly caused by the debris of Gaia-Enceladus.}
    \label{fig:pmlbcorrmap}
\end{figure}

Let us now inspect the cross-correlation between the velocity components projected on the sky. Previous work, by \cite{Iorio2019TheMerger} using a sample of RR-Lyrae stars, has shown that the Pearson correlation of $\mu_\ell$ and $\mu_b$ creates a pattern on the sky indicative of the halo being radially anisotropic, see for example their Fig.~5. Figure~\ref{fig:pmlbcorrmap} shows the mean correlation coefficient in bins on the sky for our RPM sample. The coefficient is calculated for the proper motions after correcting for the solar motion (i.e. using Eqs.~\ref{eq:vlvbsun} and the inverse of Eq.~\ref{eq:vlvb}), it is defined as
\begin{equation}
{\rm Corr}(\mu_\ell^\ast,\mu_b^\ast) = 
\frac{{\rm cov}(\mu_\ell^\ast,\mu_b^\ast)}{{\rm std}(\mu_\ell^\ast){\rm std}(\mu_b^\ast)}.
\end{equation}
The Pearson correlation ranges from [-1,1], given the Cauchy-Schwarz inequality.
Figure~\ref{fig:pmlbcorrmap} clearly shows that there exists a strong correlation between the two proper motion components. If the velocity ellipsoid of the stars in our sample were isotropic there would be no signal and if it were biased towards circular orbits it would have a very different signal. The correlation pattern that is observed is similar to that detected by \cite{Iorio2019TheMerger} and this is direct evidence that stars on radial orbits dominate the halo sample.

\subsection{Global kinematic maps}\label{sec:globkin}

Using the $(v_\ell^\ast,v_b^\ast)$ tangential velocities defined above, we can calculate pseudo-3D velocities. These are not the true 3D velocities because we assume that the line-of-sight velocities of all the stars are zero (i.e. $v^\ast_{\rm los} = 0$). This assumption is valid if the velocity distribution is centred on zero in the galactocentric rest frame. However, they will not be zero on average for local regions on the sky because of the imprint of the motion of the Local Standard of Rest around the Galactic Centre, following a $\sin{\ell} \cos{b}$ pattern. 

The equations for the pseudo-Cartesian velocities are 
\begin{subequations}
\begin{equation}
    \tilde{v}_x = -v_\ell^\ast\sin{\ell} - v_b^\ast \cos{\ell}\sin{b},
\end{equation}    
\begin{equation}
    \tilde{v}_y =  v_\ell^\ast\cos{\ell} - v_b^\ast \sin{\ell}\sin{b},
\end{equation}
\begin{equation}
    \tilde{v}_z = v_b^\ast\cos{b}.
\end{equation}
\end{subequations}
We adopt the notation $(\tilde{v}_x,\tilde{v}_y,\tilde{v}_z)$ for this set of velocities to make clear they are not the true Cartesian velocities. Subsequently, we calculate galactocentric cylindrical velocities and adopt a similar notation $(\tilde{v}_\phi,\tilde{v}_R,\tilde{v}_z)$. To obtain these coordinates, we place the Sun at $X = -8.2$ kpc \citep[which agrees well with the recently determined distance to the massive black hole in the centre of the Galaxy by \citealt{GRAVITYCollaboration2018DetectionHole}]{Mcmillan2017}.

\begin{figure}[ht]
    \centering
    \includegraphics[width=\hsize]{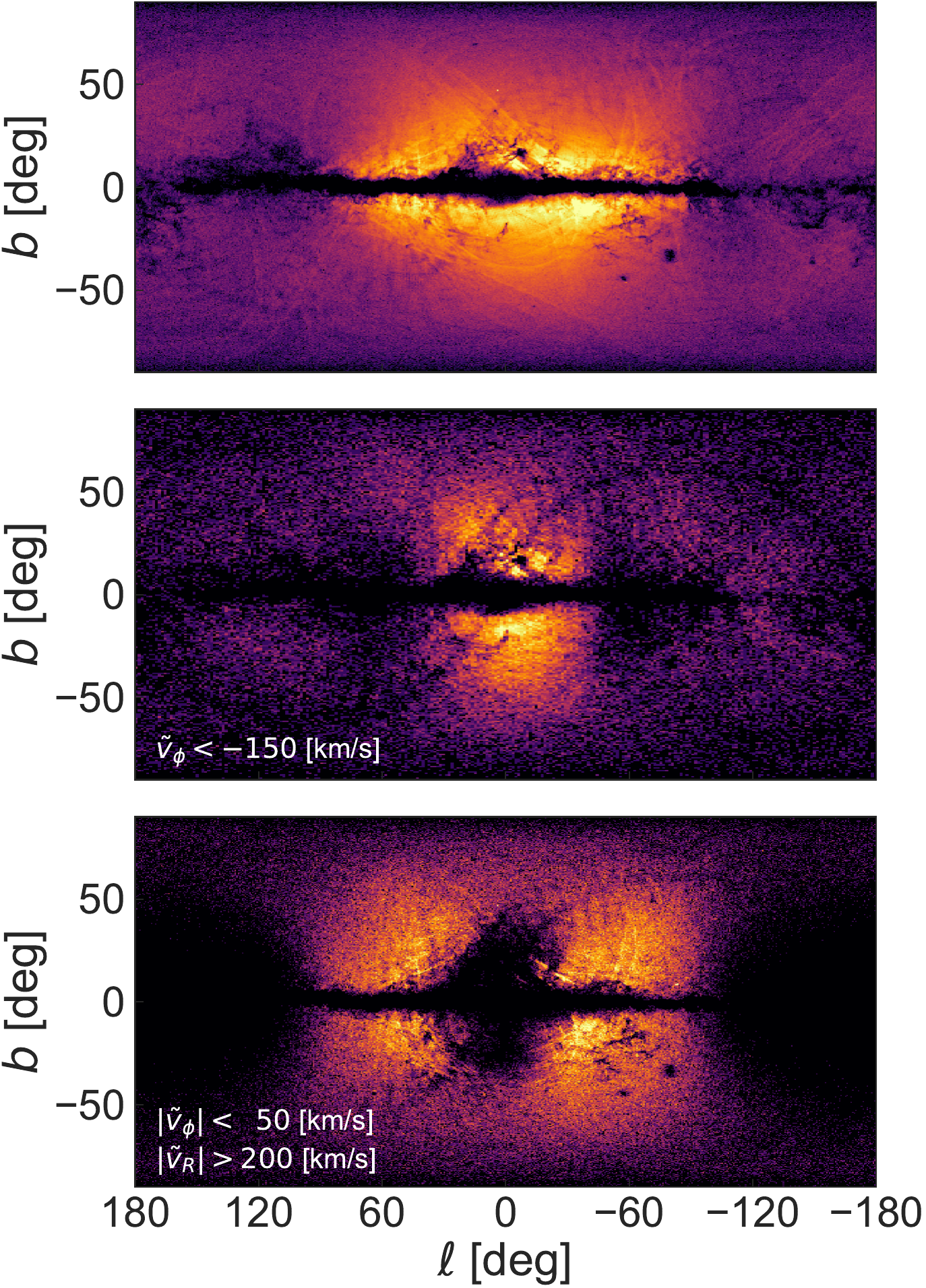}
    \caption{Distribution of halo stars projected on the sky.
    Top: All of the stars in the sample with reliable distances (see Sect.~\ref{sec:dataphotdist}).
    Middle: A subset of retrograde halo stars. 
    Bottom: A subset of halo stars on radial orbits. The footprints seen in these panels are mostly due to selection effects.}
    \label{fig:vel-skymaps}
\end{figure}
Besides looking for narrow streams, we can also use the velocity information to investigate the footprint of some interesting selections in velocity space. The two regions that we inspect are the retrograde halo, selected as all the stars with $\tilde{v}_\phi<-150 ~ {\rm km/s}$, and the radial halo, selected as $|\tilde{v}_\phi|<50 ~ {\rm km/s}$ and $|\tilde{v}_R|>200 ~ {\rm km/s}$. We do not select stars with a small pseudo space-velocity (i.e. $|\tilde{v}_R|<200 ~ {\rm km/s}$) because the missing line-of-sight velocity moves stars towards zero velocities, so this is where we expect to find significant contamination. 

Figure~\ref{fig:vel-skymaps} shows the distribution on the sky of the full sample (top), all retrograde stars (middle), and stars on radial orbits (bottom). These maps reveal a centrally concentrated halo, with most of the stars near the Galactic Centre (yellow colours). Both the retrograde maps (middle row) and radial maps (bottom row) show a clear footprint. At first sight, the footprint of the radial halo is very similar to the full-sky maps made of nearby Gaia-Enceladus stars, shown in \cite{Helmi2018} \citep[see also the skymaps presented in][]{Iorio2019TheMerger}. However, this footprint is affected by the selection effects of the RPM sample. The kinematic selections imposed in the middle and bottom panels can also impact the distribution of the stars on the sky, depending on what the intrinsic velocity distributions are. 

In Appendix~\ref{sec:appselef} we explore how the selection biases that are introduced by the incomplete velocity information affect the maps shown in Fig.~\ref{fig:vel-skymaps}. We use the {\it Gaia} 6D sample, but apply a high-tangential velocity selection criterion and set the line-of-sight velocities to zero. This analysis shows that such selections can produce the footprints that are very similar to those shown in Fig.~\ref{fig:vel-skymaps}. Therefore, the maps shown in this figure cannot be simply interpreted at face value.

\section{The velocity distribution of the local halo}\label{sec:velloc}
After highlighting the streams and structures in the distant halo, we now focus on exploring the velocity distribution of the local halo. Velocity space is very suitable to look for local streams in the form of overdensities. In small volumes, in which the (orbital) velocity gradients are small, stars with similar velocities have similar orbits. In this section we inspect only stars with a heliocentric distance smaller than 2 kpc, or even smaller volumes when indicated.

\subsection{Toomre selection}\label{sec:toomre}

The RPM sample is selected to contain a high fraction of halo stars, but we have seen in Sect.~\ref{sec:spatdist} that there is still some fraction of contamination from thick disc stars. Therefore we now impose a second selection to filter out thick disc stars based on their kinematics, namely, we remove stars that have $|{\bf \tilde{V}}-{\bf V}_{\rm LSR}|<250~{\rm km/s}$. This selection is similar to a `Toomre' cut, which is often used to differentiate disc from halo stars \citep[e.g.][]{Bonaca2017, Koppelman2018, Posti2018TheEllipsoid}. We adopt a rather strict limit of $250~{\rm km/s}$ here because the $\tilde{v}$ velocities are not the true 3D velocities. In total, we are left with $3~223~725$ high-quality halo stars ($\sim 30\%$ of the total sample). This number is in concordance with the fact that the red sequence (hot thick disc) contributes about 50\% to the sample of stars with $v_{\rm tan}>200~{\rm km/s}$ \citep[e.g.][see also \citealt{Amarante2020TheDR2}]{Sahlholdt2019CharacteristicsDR2}.

\subsection{Consistency check with RVS sample}\label{sec:RVS}
\begin{figure}
    \centering
    \includegraphics[width=\hsize]{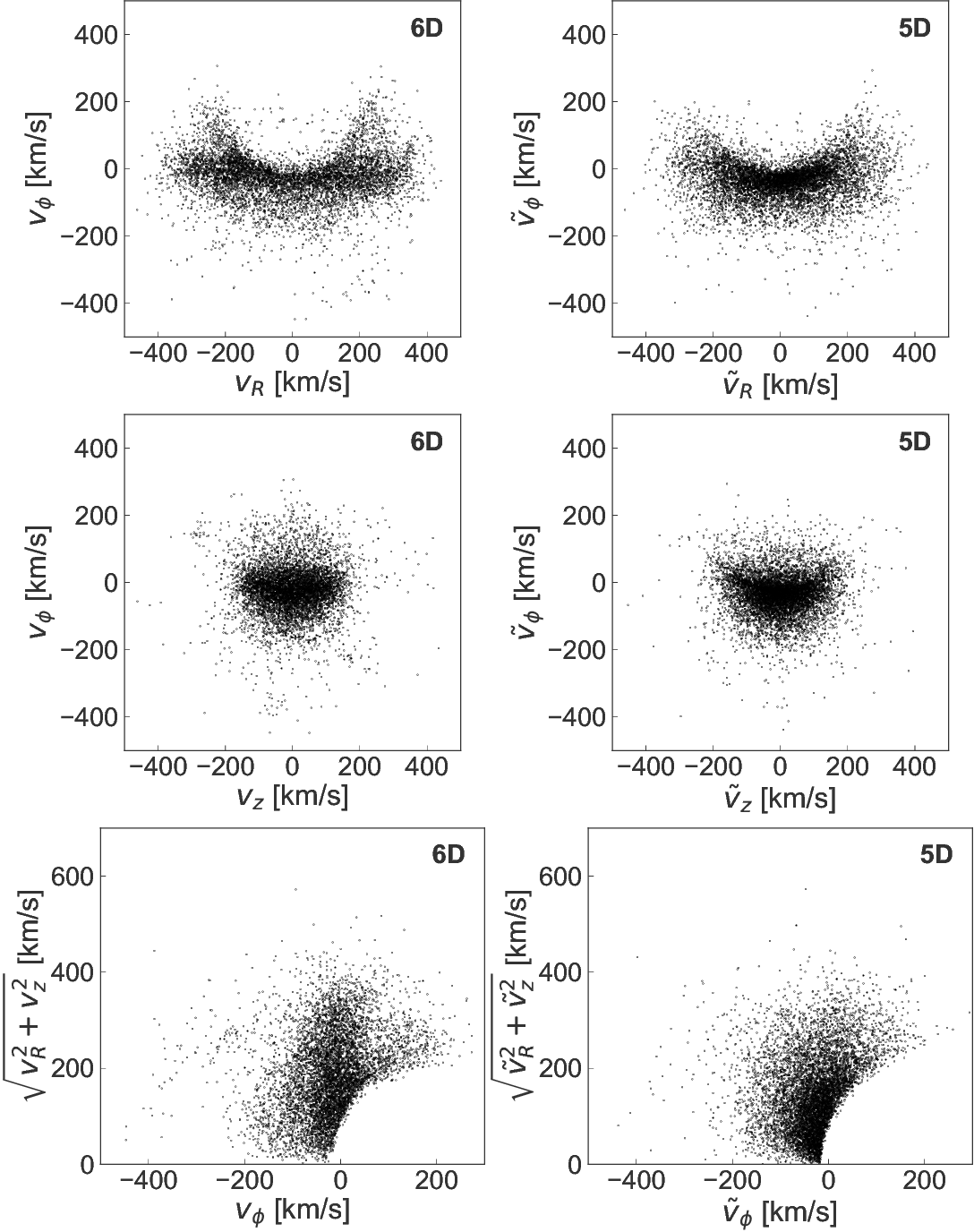}
    \caption{Velocity distribution of halo stars in the solar neighbourhood (distance $<2$ kpc) selected from the {\it Gaia} 6D sub-sample. The velocities calculated using the full phase-space information of the stars are shown in the left panels, while on the right they have been computed by setting the line-of-sight velocities to zero. Most of the structures present in the full 6D sample are strongly diluted in the 5D case.}
    \label{fig:SN-6D-2-5D}
\end{figure}

In Sect.~\ref{sec:globkin} and Appendix~\ref{sec:appselef} we have seen that the RPM selection method introduces selection effects in subsets that are selected kinematically. These features vary with location on the sky (i.e. the `blindspots'). We check if the missing line-of-sight velocities also create features in velocity space locally. For this check, we use a control sample of nearby halo stars selected from the {\it Gaia} 6D sample, which includes both giants and main sequence stars. See Appendix~\ref{sec:ApRVS} for a summary of this halo sample, with stars within $< 2$~kpc. We aim to compare the RPM selection method to one using the full phase-space information. After applying the (pseudo) Toomre selection, we find about $6700$ halo stars based 6D information and $\sim 7700$ when artificially setting the line-of-sight velocity to zero. 

Figure~\ref{fig:SN-6D-2-5D} shows the velocity distributions of halo stars in the 6D sample. Both the true velocities (left) and the pseudo-velocities (right) are shown. Only those stars with line-of-sight velocities close to zero are found in the same location left and right. The disc in these spaces is centred on $v_\phi \approx 230$ km/s, the horseshoe shape (top panel) is an artefact of its removal (the Toomre cut). Subtle structures present in the true velocity distributions are diluted in the pseudo-velocities. Even the footprint of Gaia-Enceladus \citep{Helmi2018, Belokurov2018Co-formationHalo}, which creates an elongated structure in $v_R$ at constant $v_\phi$, is blurred. For a much larger sample of stars we should be able to find faint imprints of the structures present in the true velocities, simply because by chance there will be stars with close to zero radial velocities. Comforted by the fact that no artificial structures are created by the introduction of the pseudo-velocities we turn back to the RPM sample.

\subsection{Local streams in velocity space}

\begin{figure*}
    \centering
    \includegraphics[width=\hsize]{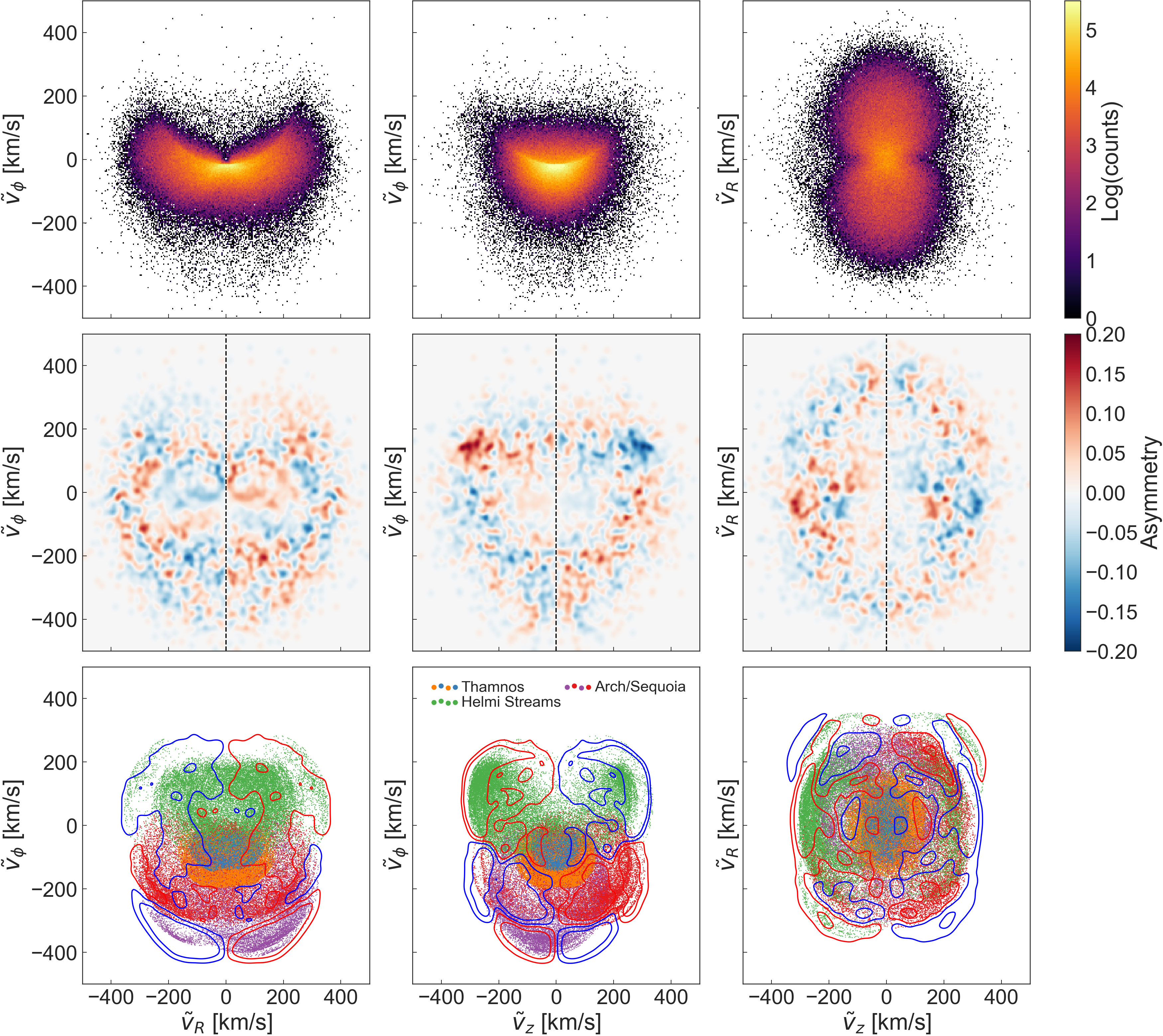}
    \caption{
    Velocity distributions of local ($d<2$ kpc) halo stars calculated without the line-of-sight component. For each combination of the cylindrical velocity components a 2D-histogram is shown in the top row, while the asymmetry (defined as in Eq.~\ref{eq:asymm}) with respect to the velocity plotted on the $x$-axis is shown in the middle row. We compare the maps to the asymmetry originating from a few structures that have been identified in the 6D sample in the bottom row.
    The asymmetry fleshes out structures that are asymmetric in either $\tilde{v}_R$ or $\tilde{v}_\phi$. Structures that are clearly present in the RPM sample, see middle row, are the Helmi Streams (green dots) and Sequoia (red and purple dots).}
    \label{fig:vel-SN}
\end{figure*}

We start with the analysis of the velocity distribution, see Fig.~\ref{fig:vel-SN}. The sample that is shown results from the RPM sample after applying the Toomre criterion. The top row shows the 2D-histogram of the velocities in $256\times256$ bins. The elongation in $\tilde{v}_R$ at constant $\tilde{v}_\phi$ of the velocity distribution is reminiscent of the footprint of Gaia-Enceladus (c.f. Fig.~\ref{fig:SN-6D-2-5D}). All three panels show relatively smooth distributions, with only subtle hints of substructures. To enhance these structures, we inspect the asymmetry of the distributions in the horizontal axis of each panel, see the middle row. We define the asymmetry as

\begin{equation}
\label{eq:asymm}
    {\rm Asymmetry} = \frac{H'_{+} - H'_{-}}{H'_{+} + H'_{-}},
\end{equation}
where $H'_{+,-}$ are smoothed histograms given by
\begin{equation}
    H'_{+,-} = \mathcal{N}(H_{+,-}+1,2).
\end{equation}
Here $H_{+}$ is the histogram as it appears in the top row, $H_{-}$ is its mirrored counterpart with the sign of $\tilde{v}_R$ or $\tilde{v}_z$ flipped and $\mathcal{N}(H,2)$ is a Gaussian filter of the histogram $H$ with kernel-size $2$. The histograms range from $-500$ km/s to $500$ km/s in both directions with $250\times250$ bins. A Gaussian kernel of $2$ corresponds to a standard deviation of $\sim 8$ km/s, which is roughly the mean velocity error of this sample. Positive values (red) indicate an overdensity and negative (blue) an under density. By construction, each overdensity is matched by a conjugate underdensity  mirrored with respect to the panel's $x=0$ axis.

The asymmetry maps reveal both small overdensities and large-scale patterns. These overdensities imply the presence of non-phase-mixed debris. Any population in the halo that is sufficiently phase-mixed will not display an asymmetry in the (local) velocity distribution. To shed light on the origin of the asymmetries, we map several structures detected in the 6D sample in the bottom row. These maps are based on the structures found by \cite{Koppelman2018}. We focus on these structures because they are asymmetric in the true velocities and therefore might also be in 5D sample. 

To establish if there is a link between the structures in the 6D and 5D samples, we first need to bear in mind the following issues:
$i$) The substructures from the 6D samples only comprise few stars each and hence they do not sample the sky densely; $ii$) The values of the pseudo-velocities $\tilde{v}$ depend on location on the sky, so we cannot
use these stars directly to predict where stars in the streams in the 5D sample would be located in velocity space; $iii$) We expect the streams to be broad enough for the member stars to be isotropically distributed on the sky. With these caveats in mind we aim to enhance the number of sources per group in the 6D sample by resampling each star using the following method:
\begin{itemize}
    \item Each of the 103 stars is re-sampled 1000 times,
    \item For each realisation, we generate a random location on a sphere of 2 kpc in radius that is centred on the solar position (uniformly distributed on the surface),
    \item We assign the random location with the unchanged velocity vector (in galactocentric coordinates),
    \item We calculate the $\tilde{v}_i$ velocities for each star based on its new location.
\end{itemize}
We assume here that the Galactic potential does not vary over the volume in which we resample the structures. In this approximation, the orbits of the stars are mostly determined by the amplitude and direction of the velocities since to first order the location-dependent potential term is constant. The resampling is thus a simple way of modelling more members of the stream. That is, to add stars on the same orbits as the detected streams.

This resampling shows what the footprint of the structures identified in the 6D sample could look like in the pseudo velocity-space. The stars are coloured by the original labels of \cite{Koppelman2018}. The structure indicated with green dots has been associated with the Helmi Streams \citep[e.g.][]{Helmi1999,Koppelman2019a}. Recent studies suggest that the structures indicated with red and purple dots are part of a structure in velocity space known as Sequoia \citep{Myeong2019EvidenceHalo}, and that the blue and orange structures are part of a structures labelled Thamnos \citep{Koppelman2019}.

We overlay the expected asymmetry over the newly sampled members of the structures identified in the 6D set (blue and red lines in the bottom row). These asymmetry contours do a surprisingly good job in explaining the asymmetries seen in the 5D sample. Therefore, we conclude that the features that are shown in the middle row for this sample are mainly due to the structures previously detected in the {\it Gaia} sample with full phase-space information.

Apart from these structures, there are several other overdensities visible of a few km/s in size (small red and blue dots). To check the statistical significance of the asymmetries found we shuffle the data. We randomly shuffle the velocities of Fig.~\ref{fig:vel-SN} (top panel) and for each random set we create an asymmetry map. The Helmi Streams and a handful of other groups (those with the darkest colours) have a strong asymmetry. However, similar levels of asymmetry are found in the random realisations of the data. We note that this is a very crude estimate of the statistical significance as only the amplitude and not the extent of the asymmetries are taken into account. For example, Sequoia (purple and red dots in the bottom panel) overlaps with several of the asymmetries seen in the middle row. It spans most of the retrograde part of the diagram ($\tilde{v}_\phi<-100~{\rm km/s}$). Therefore, we surmise that this asymmetry is due to the debris of Sequoia \citep{Myeong2019EvidenceHalo} and possibly also Thamnos \citep{Koppelman2019}. Of course, the asymmetries seen in Fig.~\ref{fig:vel-SN} can also be partly due to unidentified halo structures in velocity space.

We perform an additional test to measure the overall level of asymmetry in the dataset (rather than of a given feature). We compute the `total level of asymmetry' per map as the sum over all the bins of the absolute values of the asymmetries per bin both for the data as well as for 1000 randomly reshuffled samples. For these samples, we compute the average and its dispersion $\sigma$. Finally we measure a significance value as $({\rm Total Asymmetry}_{\rm data} - \langle {\rm Total Asymmetry}\rangle_{\rm rand}) /\sigma$. We find that the significance levels vary per map (i.e. combination of cylindrical velocities), taking values $10.2\sigma$, $12.5\sigma$, and $5.8\sigma$ from left to right in Fig.~\ref{fig:vel-SN}. These results imply that, while individual asymmetries might be caused by random clustering of the stars, the total level of asymmetry found in the data is far from random.

\subsection{Excess of pairs}

\begin{figure}
    \centering
    \includegraphics[width=\hsize]{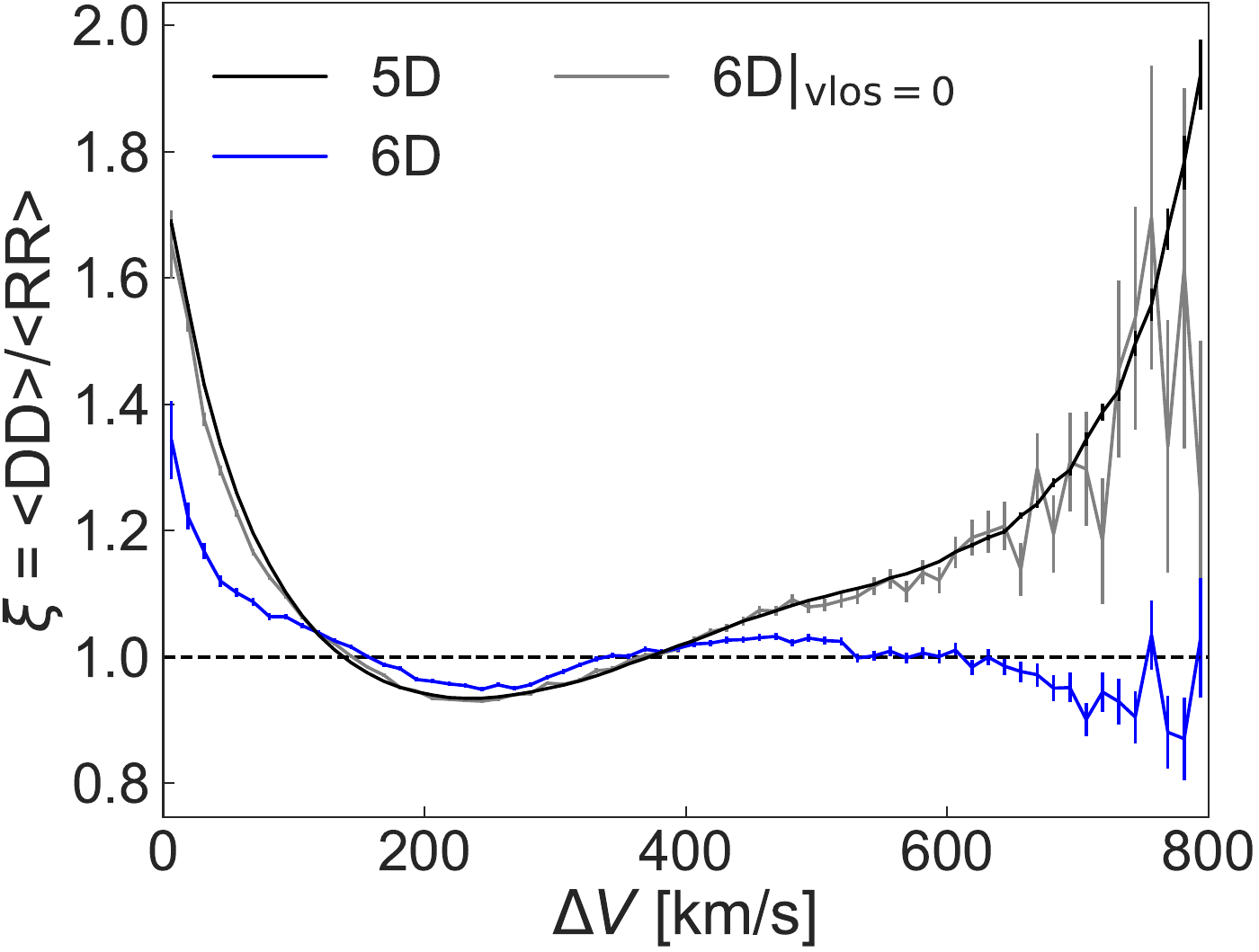}
    \caption{Two point correlation function for stars in a local volume ($d<1$ kpc). Both the RPM sample (black) and the 6D (blue) control sample are shown, in addition to a 5D version of the latter (grey). There is an excess of stars at velocity scales smaller than 100 km/s, and at scales larger than 400 km/s in the 5D case. The correlation in the RPM sample is fully compatible with the correlation in the 5D version of the control sample.}
    \label{fig:velcorfun}
\end{figure}

As final part of this work, we quantify the clustering of stars in velocity space with a two-point correlation function. That is, we check if the stars are randomly distributed in velocity space, or whether they tend to cluster on some velocity scale. We use a correlation function of the form
\begin{equation}
    \xi(\Delta v) = \frac{DD (\Delta v)}{RR (\Delta v)},
\end{equation}
where $DD (\Delta v)$ is the number of pairs with a (pseudo) 3D velocity difference of $\Delta v$ and, similarly, $RR (\Delta v)$ is the number of pairs in a randomised sample. We obtain the randomised sample by shuffling the velocities and we count the pairs in 64 bins ranging from $0$ to $800$ km/s. For computational reasons, we only calculate the correlation function for stars inside a volume of 1 kpc.

Figure~\ref{fig:velcorfun} shows the velocity correlation function $\xi$ for the RPM sample (black) and for a control sample of stars with full phase-space information (blue). Similar to analysis presented in Sect.~\ref{sec:RVS}, we show the effect of the missing line-of-sight velocities with the control sample (grey curve). The error bars indicate the error in $\xi$, calculated as the Poisson error in the number of pairs. A correlation of $\xi>1$ indicates grouping of stars in excess of random clustering. All curves show significant clustering on velocity scales $\Delta v \lesssim 100~{\rm km/s}$. Clustering of velocities on small scales hints at the presence of streams.

The correlation excess at large velocity scales is less intuitive to interpret. However, a similar effect is seen in the 5D version of the 6D control sample. The grey and black curve are strikingly similar. Thus, the missing velocity component artificially enhances the correlation of the velocities on large scales and to a lesser extent also on small scales.

It is interesting that the black and grey curves match so well. This must mean that in spite of the roughly 10 times smaller size of the 6D sample (2945 stars), compared to the RPM sample, in both cases, most of the local streams are well resolved and contain at least a few stars per stream. This would be consistent with the early predictions by \citep[e.g.][]{Helmi1999a} who estimated of the order of 300 - 500 streams should be present near the Sun for a halo with a pure accretion origin. The advantage of a larger sample is that the amplitude of the correlation will be measured more accurately, as the (Poisson) uncertainty decreases with sample size.

Recently, \cite{Simpson2019SimulatingNeighbourhood} have investigated the velocity correlation function for Milky Way-like galaxies in the Aurigaia mock catalogues \citep{Grand2018Aurigaia:Simulations}. Several of the halos analysed by these authors show similar correlation functions, both on small and large scales, to the one observed in the RPM sample. An excess of pairs can be very difficult to interpret if there are too few particles, also if the volume is not localised. Pairs always appear as a consequence of substructure and this can be accreted but also {\it in situ} (mergers do produce features in the {\it in situ} population as well, e.g. \cite{Gomez2012SignaturesRevisited,Jean-Baptiste2017OnTale}).
\section{Discussion and conclusions}\label{sec:conclusions}

We have explored the use of a reduced proper motion (RPM) diagram to identify tentative main sequence (MS) halo stars in {\it Gaia} DR2. This method makes only use of the {\it Gaia} photometry and proper motions. Most conventional methods rely on distance information or spectroscopic observations of the stars, which limits the sample size substantially. With the RPM selection method we find $11~711~399$ tentative halo MS stars.

For these stars, we calculate photometric distances using the relatively simple colour-magnitude relation of the main sequence in the {\it Gaia} photometric bands. These distances have typical errors of $\sim 7$\%, which makes them much more reliable than inverted parallaxes. The distances are especially accurate for stars with colours $0.45 < G-G_{\rm RP} < 0.715$. Stars with bluer colours are near the MS turn-off and their magnitudes are a very steep function of colour. On the other hand, the MS broadens significantly for redder colours, which causes the error in the photometric distance calibration to increase. The above-mentioned colour range reduces our sample to $7~117~555$ MS halo stars with good distances. The median velocity errors for these stars are $\sim 20$ km/s.

A limitation of dealing with a sample of halo MS stars is that they are intrinsically faint. The most distant MS star in our particular sample is found at $\sim16$ kpc. A star with $M_G\approx5$ (the brightest given the colour cut) at this distance would have the limiting magnitude of the {\it Gaia} DR2 catalogue.  Beyond a heliocentric distance of $\sim 5$ kpc, the sample suffers from incompleteness, and thus halo overdensities such as the Virgo and Hercules-Aquila clouds cannot be studied with our dataset. 

The spatial distribution of the RPM sample displays large-scale patterns and structures but it is not trivial to disentangle real structure from artefacts of the selection method. Nonetheless, full-sky maps of the mean velocities $v_\ell$ and $v_b$, reveal the presence of some known streams, such as GD-1 \citep{Grillmair2006DETECTIONSURVEY}. Since pixels on the sky that overlap with a stream may have a mean velocity that is different from that of the background, and a smaller velocity dispersion, this promises to be an interesting approach to identifying substructures. Future {\it Gaia} data releases will be less affected by systematics due to, for example, the scanning-pattern, and will provide more precise proper motions leading to easier distinction between stream and background stars.

Another promising approach is to calculate (pseudo) space velocities of the stars by assuming their line-of-sight velocity to be zero. Most of the structure in velocity space is smoothed out because of the missing velocity component. However, by chance, some stars can have a true line-of-sight velocity of zero, particularly if the sample is large enough. In that case, we may still find some structure in velocity space.

In fact, in pseudo-velocity space for nearby MS halo stars we find clear imprints of the Helmi Streams. Also the footprint of Gaia-Enceladus (i.e. the Gaia-Sausage) is found in several of the velocity maps that we explore. Moreover, the retrograde halo shows a strong asymmetry in the velocity distribution, reminiscent of the accreted structures that have been previously reported in the 6D sample \citep{Koppelman2018, Koppelman2019, Myeong2019EvidenceHalo}. The total level of non-phased-mixed substructure in pseudo-velocity space, as measured by an asymmetry parameter, is very significant ($\gtrsim 6\sigma$), in comparison to randomised samples.

Through a two-point velocity correlation function we measure a very significant excess of stars with small velocity differences ($\Delta v <100~{\rm km/s}$). The amplitude of this clustering signal for the RPM sample is similar to that obtained when using 6D  {\it Gaia} sample, implying that this sample is already large enough for a true quantification of the amount of clustering. This appears to be consistent with the expectation from theoretical models that hundreds of debris streams are crossing the solar vicinity \citep[e.g.][]{Helmi1999a}, as only if the sample is large enough will the streams be populated by enough stars to produce a signal. It will be particularly interesting to study this excess of close velocity pairs in more detail and, in particular, to do spectroscopic follow-up of the stars in pairs as these likely originate from very localised regions in the phase-space of accreted galaxies.

The RPM catalogue we have built provides interesting targets for spectroscopic follow-up, for example for chemical tagging. Even low/intermediate resolution spectroscopy would be highly valuable because it would provide the missing line-of-sight velocity but also because even a metallicity and [$\alpha$/Fe] abundance are extremely useful to disentangle merger events from one another and to construct basic chemical enrichment histories. Halo main sequence stars are easy to identify as shown here. Moreover, they have the advantage of being very numerous and that the elements found in their atmospheres are directly representative of the elements in their birth material \cite[e.g.][]{Tolstoy2009Star-FormationGroup}, which is not necessarily true for giants, which might have undergone mixing. 

Finally, a sample such as the one presented here could be used to map the (local) mass-distribution of the Milky Way, the density profile of the stellar halo as well as its dynamical properties. In all applications it is important to bear in mind that our sample is kinematically biased by construction, and that it misses halo stars with small proper motions and large line of sight velocities.

\begin{acknowledgements}
We thank Eduardo Balbinot for helpful discussions regarding the methodology presented in this paper and for his comments on earlier versions of the manuscript. We also thank the anonymous referee for their constructive comments that helped improve the text. We gratefully acknowledge financial support from a VICI grant from the Netherlands Organisation for Scientific Research (NWO) and a Spinoza prize. This work has made use of data from the European Space Agency (ESA) mission Gaia ({\tt http://www.cosmos.esa.int/gaia}), processed by the Gaia Data Processing and Analysis Consortium (DPAC, {\tt http://www.cosmos.esa.int/web/gaia/dpac/consortium}). Funding for the DPAC has been provided by national institutions, in particular the institutions participating in the Gaia Multilateral Agreement. 
For the analysis, the following software packages have been used: {\tt vaex} \citep{Breddels2018}, {\tt numpy} \citep{Oliphant2006GuideNumPy, VanDerWalt2011TheComputation}, {\tt matplotlib} \citep{Hunter2007Matplotlib:Environment}
, {\tt jupyter notebooks} \citep{Kluyver2016JupyterWorkflows}.
\end{acknowledgements}

\bibliographystyle{aa} 
\bibliography{AMainFile} 

\appendix

\section{Velocity maps without MSTO stars}\label{sec:Apvelmaps}

\begin{figure*}
    \centering
    \includegraphics[width=\hsize]{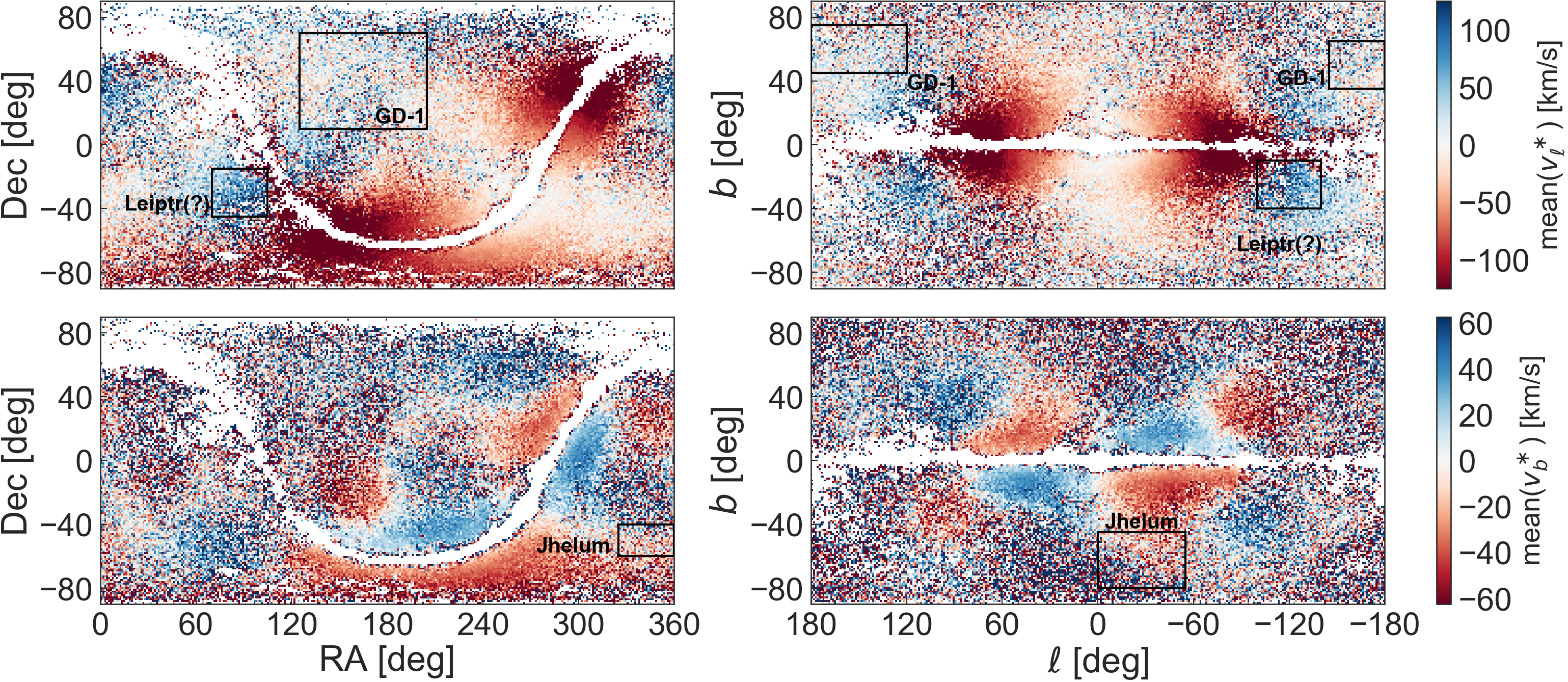}
    \caption{Same as Fig.~\ref{fig:vlvb}, but without MSTO stars.}
    \label{fig:old-vlvb}
\end{figure*}

In Sect.~\ref{sec:velmom} we show the velocity moments of the RPM sample binned on the sky. To flesh out distant streams, such as GD-1, we have included sources that are near the MSTO. However, the MSTO stars might have distances that are underestimated by up to $40\%$. For this reason, we show here the same figures, but without these MSTO stars, see Fig~\ref{fig:old-vlvb}
(and Fig.~\ref{fig:vlvb}
for comparison). Clearly, the overall contours are the same with and without the MSTO stars. However, the figures shown in this appendix do not show any (clear) narrow streams.

\section{Selection effects} \label{sec:appselef}

\begin{figure}
    \centering
    \includegraphics[width=\hsize]{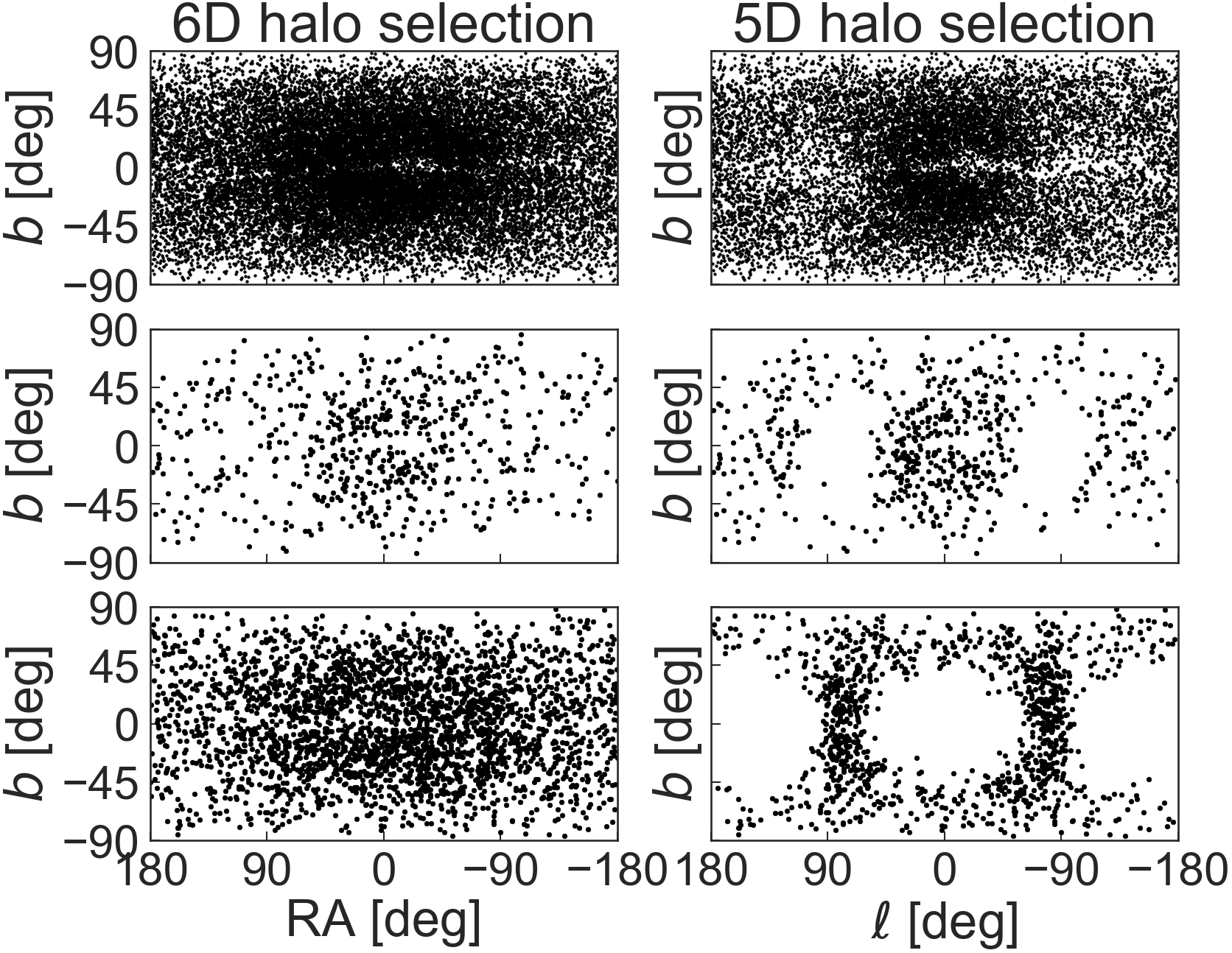}
    \caption{Selection effects of the RPM halo-selection, illustrated with a sample of halo stars for which the full phase-space information is available (see Appendix~\ref{sec:ApRVS}). The layout of the panels is based on Fig.~\ref{fig:vel-skymaps}. The left columns shows the distribution on the sky of halo stars selected using full phase-space information. On the right, we show the distribution of a halo sample selected without the line-of-sight velocity.}
    \label{fig:selectioneffect-6D5D}
\end{figure}

If, for simplicity, we assume that all of the halo stars are on radial orbits, then there would be blind spots in the RPM sample towards the Galactic Centre and anti-centre. In these directions, stars on purely radial orbits move along the line-of-sight. Therefore, their proper motions are close to zero, and they will not be present in the RPM selection. The blind spot towards the Galactic Centre is smaller than its conjugate spot towards the anti-centre. 

In Fig.~\ref{fig:selectioneffect-6D5D} we investigate this selection effect by showing the sky-distribution of halo stars in the 6D sample. We show a sample of halo stars selected using the full phase-space information that is available (left column) and a sample of halo stars selected using only tangential velocities (right column). The sample that is used is identical to the one described in Appendix~\ref{sec:ApRVS}. The layout (per column) is the same as the layout of Fig.~\ref{fig:vel-skymaps}. The similarity between the footprints of the red dots and the distribution displayed in Fig.~\ref{fig:vel-skymaps} are striking. We have to conclude that the large structures that are seen in Fig.~\ref{fig:vel-skymaps} are dominated by selection effects. 

We do note that the distance distribution of the stars in the full phase-space sample is more local. Typically these stars are within 3 kpc, while the RPM sample has a median distance of 4.39 kpc. Therefore, the maps of Fig.~\ref{fig:selectioneffect-6D5D} might look different when a larger (volume-wise) sample is used. Although, up to first order they should be the same. These maps mostly show where the motion aligns the most with the line-of-sight, which is a sky-dependent phenomenon and not distance dependent.

\section{RVS sample}\label{sec:ApRVS}

In Sect.~\ref{sec:RVS} we have used the RVS \citep{Katz2019GaiaVelocities} subset of {\it Gaia} DR2 as a control sample. This sample of stars comprises full phase-space information. 
For the construction of this sample we start with the full RVS sample and impose the following quality criteria: Firstly, ${\tt phot\_g\_mean\_flux\_over\_error} > 50$ and ${\tt phot\_rp\_mean\_flux\_over\_error} > 20 $, secondly ${\tt visibility\_periods\_used} >=5 $, and finally ${\tt parallax\_over\_error} > 5$.


We calculate Cartesian coordinates and velocities similarly as described in Sect.~\ref{sec:veldist} and Sect.~\ref{sec:velloc}, as well as tangential velocities according to Eq.\eqref{eq:vtan}. We then select a halo sample as stars with $v_{\rm tan} > 200~{\rm km/s}$. Because for this set there are line-of-sight velocities available, we also calculate the true and pseudo-velocities, and then impose the following cut in velocity space:
\begin{equation}
    |\mathbf{V} - \mathbf{V}_{\rm LSR}| > 250~{\rm km/s}.
\end{equation}
We calculate this cut separately for the true and pseudo-velocities. As a result we end up with two halo samples, one based on full phase-space information and one without line-of-sight velocities.
For a local sample ($<2$ kpc), we find 6718 (using true velocities) and 7749 (using pseudo-velocities) tentative halo stars.

\end{document}